\documentclass[astrosymb]{aastex631}

\shorttitle{M87 lepto-hadronic model}
\shortauthors{Boughelilba et al.}

\graphicspath{{./}{figures/}}

\begin{document}

\title{Lepto-hadronic jet-disc model for the multi-wavelength SED of M87}

\correspondingauthor{Margot Boughelilba}
\email{margot.boughelilba@uibk.ac.at}

\author[ 0000-0003-1046-1647 ]{Margot Boughelilba}
\affiliation{Institute for Astro and Particle Physics, University of Innsbruck, 6020 Innsbruck, Austria}
 
\author[0000-0001-8604-7077]{Anita Reimer}
\affiliation{Institute for Astro and Particle Physics, University of Innsbruck, 6020 Innsbruck, Austria}

\author[0000-0003-1332-9895]{Lukas Merten}
\affiliation{Ruhr-Universität Bochum, Institut für Theoretische Physik IV, 44801 Bochum, Germany}
\affiliation{Institute for Astro and Particle Physics, University of Innsbruck, 6020 Innsbruck, Austria}

\begin{abstract}

The low-luminosity Active Galactic Nuclei M87, archetype of Fanaroff-Riley I radio-galaxies, was observed in a historically quiet state in 2017. While one-zone leptonic jet models alone cannot explain the core radio-to-gamma-ray spectrum, we explore a hybrid jet-disc scenario. In this work, we model the overall spectral energy distribution of M87's core with a dominating one-zone lepto-hadronic jet component, coupled with the contribution from the accretion flow. We find close-to-equipartition parameter sets for which the jet component fits the radio-to-optical data as well as the gamma-ray band, while the accretion flow mainly contributes to the X-ray band. The effects of gamma-ray absorption by the Extragalactic Background Light during the propagation towards Earth are probed and are found to be negligible for this model. The neutrino flux produced by such scenarios is also calculated, but remains below the current instruments sensitivity.
\end{abstract}


\section{Introduction} \label{sec:intro}

M87 is one of the closest examples 
of low-luminosity Active Galactic Nuclei (AGN), located at a distance of $\sim 16.8$ Mpc from Earth (corresponding to a redshift $z \approx 0.004$) in the Virgo cluster. The mass of the supermassive black-hole at its center was estimated around $6.5 \times 10^9 \mathrm{M}_\sun$ \citep{2017_Mass_EHT}.
In 2017, an extensive multi-wavelength observation campaign was launched, taking quasi-simultaneous data from several telescopes over the entire electromagnetic band \citep{EHTpaper}. For nearly 2 months, M87's core region was observed in a particularly low state. These observations allow to study the innermost radiation from the AGN, in particular the launching region of the jet, that M87 exhibits, as the broadband spectrum of these observations is dominated by the emission from the core and not from the jet's knots such as HST-1.
Furthermore, the closeness and the size of M87 make it a prime candidate accelerator of the observed high-energy cosmic rays (see, e.g. \citealt{UHECR_M87, M87UHECR_TeV}).
To explain the multi-wavelength spectral energy distribution (SED) of M87, different emission models are usually probed. Leptonic jet models are, for the case of M87, typically based on the Synchrotron-Self-Compton mechanism: synchrotron photons produced by the interaction between the jet's relativistic electrons and positrons with the ambient magnetic field are used as a target field for inverse Compton scattering by the same particles, thereby producing high-energy radiation. In \cite{EHTpaper}, two different one-zone leptonic models were applied, but failed to reproduce both, the high and low energy parts of the SED of M87 at the same time.

On the other hand, lepto-hadronic models have been proposed to explain the SED of objects such as M87 (e.g. \citealt{AnitaM87}). In such models, accelerated protons are also present in the jet together with electrons, and the high-energy part of the SED is assumed to be the result of proton-initiated processes. A clear observational signature between these two kinds of models is the production of neutrinos in the case of lepto-hadronic models.
In this paper, we explore a global model coupling the jet lepto-hadronic emission and the accretion flow, in order to explain the observed SED of M87. With this model, all the emission would originate from the core region of the AGN. 

The paper is organized as follows: in Section \ref{sec:jet section} we describe the jet model, then in Section \ref{sec:adaf} we detail the accretion flow model component and its parameters. Results of the simulations are presented in Section \ref{sec:results} and we conclude about the global core emission in Section \ref{sec:conclusion}.

\section{Jet model component} \label{sec:jet section}

As of today, the one-sided jet launching from M87 has been well-studied in all different wavelengths. In the 2017 observation campaign, \cite{EHTpaper} did not infer any time variability in the flux above 350 GeV. The data collected focus on the core emission. The angular resolution of radio observations suggests that the radio emission region is close to the jet launching region. While launch mechanisms are still unclear, the total estimated jet power for M87 of $\sim 10^{43-44} \, \mathrm{erg}\, \mathrm{s}^{-1}$ \citep{Prieto_M87, Jet_power, Jet_power_Stawarz} can be provided through, e.g. the Blandford-Znajek mechanism \citep{Blandford_Znajek, EHT2019_BZ_BP}.


In this paper we explore models that can reproduce the quiet and steady state of M87's core observed between March and May 2017. There is evidence of sub- to superluminal motion of radiating jet components in M87's inner jet (eg Snios et al 2019, Walker et al 2018), that can support a jet model setting in which the emission region is viewed as a moving blob. On the other side, the possibility that the jet is a continuous zone in which the particles flow is often considered (see, e.g. \citealt{BlandfordSteady, SteadyReview}) in the case of quiet state emission, and cannot be ruled out. We investigate both scenarios here. 
First we consider the jet 
emission region as a spherical blob with a constant radius $r'_\mathrm{em}$ of magnetized plasma moving at a mildly relativistic speed along the axis of a, during the observation time, non-expanding jet, inclined by an angle $\theta$ with respect to the line of sight. This defines a Doppler factor $\delta_\mathrm{j} = \Gamma_\mathrm{j}^{-1}(1 - \beta_\mathrm{j}\cos{\theta})^{-1}$ where $\Gamma_\mathrm{j}$ and $\beta_\mathrm{j}c$ are the bulk Lorentz factor and velocity respectively. 
For the second scenario we consider the jet as a continuous cylinder of radius $r'_\mathrm{em}$ and proper length $l' = \Gamma_\mathrm{j} l$, with $l$ being the observed length.

The EHT observation provides a strong constraint on the size of the emission region, as the angular resolution allows to probe the closest regions to the black hole. At 230 GHz, the radio flux was measured with an angular resolution $\theta_\mathrm{obs}$ of 0.06", corresponding to 7.5 $r_\mathrm{g}$ (for M87, $r_\mathrm{g} = G M_\mathrm{BH}/c^2 \approx 9.8 \times 10^{14} \, \mathrm{cm}$) in radius. However, even for a mildly relativistic jet velocity, the blob travels farther than 7.5 $r_\mathrm{g}$ over the observation time. In the continuous jet scenario, we assume that the jet is launched around the innermost stable orbit of the black hole, i.e. $6 \, r_g$ for a static black hole. Hence, when observing the core region within $7.5\, r_g$ at $230 \, \mathrm{GHz}$ the jet component is likely not the dominant one. Since we choose to focus on the core emission, we take care that for an emission region of size $\lesssim 7 \, r_\mathrm{g}$, the predicted radio flux does not exceed this particular data point.  
 %

Furthermore, the SED of M87 indicates a self-absorbed, stratified jet below at least 86 GHz \citep{EHTpaper, Blandford_prediction_SSA}. This lower limit on the self-absorption frequency $\nu_\mathrm{ssa, obs}$ and corresponding flux $S_{\nu_\mathrm{ssa, obs}}$ coupled with the estimate of the size of the emission region allows to derive a relation for the magnetic field strength B required. Following the treatment by \cite{Kino_SSA} for a moving blob:
\begin{eqnarray}
B &=& b(p_e)\left(\frac{\nu_\mathrm{ssa,obs}}{1\mathrm{GHz}}\right)^5\left(\frac{\theta_\mathrm{obs}}{1\mathrm{mas}}\right)^4\left(\frac{S_{\nu_\mathrm{ssa},\mathrm{obs}}}{1\mathrm{Jy}}\right)^{-2}\left(\frac{\delta}{1+z}\right) \quad \mathrm{G} 
\end{eqnarray}
with $b(p)$ described in the Appendix \ref{b(p)}.
From this, we estimate an order of magnitude for the magnetic field strength and then adjust the primary electron injection parameter so that the synchrotron radiation produced is of the order of $S_{\nu_\mathrm{ssa},\mathrm{obs}}$ at the given frequency. Considering that the self-absorption frequency is $\lesssim 230 \mathrm{GHz}$ (around the EHT data point), with the observed flux being $S_{\nu_\mathrm{ssa},\mathrm{obs}}\sim 0.6 \, \mathrm{Jy}$, gives an estimate for the magnetic field strength between $\sim 5 - 60$ G.

We assume that the emission region contains primary relativistic electrons and protons that are isotropically and homogeneously distributed in the comoving jet frame, and following a power-law energy spectrum cutting off exponentially, such that the spectral number density $n'_{e,p}(E') \propto E'^{-p_{e,p}}e^{-E'/E'_{\mathrm{max},e,p}}$ cm$^{-3}$, for $E' \ge E'_{\mathrm{min},e,p}$ (where e,p denotes the electrons or the protons, respectively).

These primary particles are injected continuously into the emission region at a rate $q_i$ (cm$^{-3}$s$^{-1}$), where they suffer from different interactions. These are photo-meson production, Bethe-Heitler pair-production, inverse-Compton scattering, $\gamma$-$\gamma$ pair production, decay of all unstable particles, synchrotron radiation (from electrons and positrons, protons, and $\pi^\pm$, $\mu^\pm$ and $K^\pm$ before their respective decays) and particle escape. Positrons are treated the same way as electrons, hence in the following we will use electrons to refer to the two populations irrespective of their type. 

Primary particles can also interact with external target photon fields (i.e. produced outside the jet). However, no evidence of a dusty torus has been found \citep{No_DT_in_M87} and no Fe K$\alpha$ line has been observed to support the existence of a strong broad-line region (BLR) component \citep{DiMatteoM87}. This is in line with the properties of "true" type 2 AGN (\citealt{LLAGN_NLR}; or see \citealt{Review_DT_BLR} for a review). Hence we do not consider the dusty torus nor the BLR as external target fields. On the other hand, the accretion flow could serve as an external photon field for the jet's primary particles, a possibility we discuss in Section \ref{sec:results}. 
 

The maximum energy of the primary particles is determined by $E_\mathrm{max} = \min{(E_\mathrm{max}^\mathrm{Hillas}, E_\mathrm{max}^\mathrm{loss/acc})}$ where $E_\mathrm{max}^\mathrm{Hillas}$ is the energy given by the Hillas criterion \citep{Hillas} and $E_\mathrm{max}^\mathrm{loss/acc}$ is the energy obtained by balancing the particles' acceleration and loss rates.

The Hillas criterion constrains the Larmor radius of the particles to be smaller or equal to the size of their acceleration region, leading to an estimate of the maximum particle energy $E_\mathrm{max}^\mathrm{Hillas} \approx 10^{21} Z\beta \left(R/\mathrm{pc}\right) \left(B/\mathrm{G}\right) \mathrm{eV}$. 



Expressions for $E_\mathrm{p,max}^\mathrm{loss/acc}$ and $E_\mathrm{e,max}^\mathrm{loss/acc}$ are obtained by equating the acceleration timescale $t_\mathrm{acc}(E_\mathrm{p,e,max}^\mathrm{loss/acc})$ and the loss timescale $t_\mathrm{cool}(E_\mathrm{p,e,max}^\mathrm{loss/acc})$ respectively. We follow the work of \cite{AnitaM87} to verify that the ratio of the two maximum energies $E_\mathrm{p,max}^\mathrm{loss}/E_\mathrm{e,max}^\mathrm{loss} = \left(m_p/m_e\right)^{4/(3-\beta)}$ is obtainable with a realistic turbulence spectrum. For, e.g. Kolmogorov diffusion, $\beta=5/3$, we get  $E_\mathrm{p,max}^\mathrm{loss}/E_\mathrm{e,max}^\mathrm{loss} \sim 6\times 10^9$. Bohm diffusion, where the magnetic field is fully tangled,  corresponds to $\beta=1$, and in the case of strong magnetic fields, Kraichnan turbulence $\beta = 3/2$ can be present \citep{Kraichnan_turbulence}.


To compute the time-dependent direct emission and cascade component from the jet's particles, we use a particle and radiation transport code (see, e.g. \cite{Anita_matrix_intro}) that is based on the matrix multiplication method described in \cite{Protheroe_Stanev_matrix} and \cite{Protheroe_Johnson_matrix}. The interaction rates and secondary particles and photons yields are calculated by Monte Carlo event generator simulations (except for synchrotron radiation, for which they are calculated semi-analytically). These are then used to create transfer matrices, that describe how each particle spectrum will change after a given timestep $\delta t$. To ensure numerical stability, we set $\delta t$ equal to the smallest interaction time for any given simulation. In each timestep, energy conservation is verified. For steady-state spectra, we run the simulation until we reach convergence, which we define here as the ratio $R_\mathrm{conv}$ between the flux at a simulation time $t$ and the flux at a simulation time $t - \delta t$. Convergence is reached when $R_\mathrm{conv} = F_\nu(t + \delta t)/F_\nu(t) < 1 \pm 10^{-3}$ .

All the calculations listed above are done in the jet frame. The observed spectrum $\nu F_{\nu}$ is then given by the frame transformation $F_{\nu} = (1+z)$$g_\mathrm{boost}$$L'_{\nu}/(4\pi d_\mathrm{L}^2)$ where $L'_{\nu}$ is the comoving luminosity from the jet with $d_\mathrm{L} = 16.8 \, \mathrm{Mpc}$ the luminosity distance of the source and $\nu_\mathrm{obs} = \delta_\mathrm{j}\nu'/(1+z)$. The Doppler enhancement factor is $g_\mathrm{boost} = \delta_\mathrm{j}^3$ for a moving blob  and $g_\mathrm{boost} = \delta_\mathrm{j}^2/\Gamma_\mathrm{j}$ for a continuous jet \citep{dopplerSikora, DopplerStawarz}. For a given comoving energy density, we obtain the intrinsic luminosity through $u'_{\nu} = (r'_\mathrm{em}/c)(L'_{\nu}/V')$, where $V'$ is the comoving volume of the emission region (i.e. depending on the geometry). We find that we can obtain the same observed flux for both jet configurations by setting the length of the continuous cylinder to $l' = 2\delta_\mathrm{j}\Gamma_\mathrm{j}r'_\mathrm{em}/3$. We apply this for the remaining part of this work, hence the results that we show in Section \ref{sec:results} are identical for the moving blob and the continuous jet scenario, given this condition.
The effect of gamma-ray absorption by the Extragalactic Background Light (EBL)  on the escaping photon beam travelling from the source to Earth is taken into account. Three different models, using different approaches to calculate the EBL SED as a function of the redshift, are used here to compute the flux attenuation factor. We use the models of \cite{EBL_franceschini}, \cite{EBL_dominguez}, and \cite{EBL_gilmore}, which are based on existing galaxy populations and extrapolates them back in time, based on the evolution of galaxy populations directly observed over the range of redshifts that contribute the most significantly to the EBL, and based on forward evolution of galaxy populations starting with cosmological initial conditions, respectively. The high-energy flux of M87 can be used to probe these models, and constrain the EBL density, especially in the far infrared band, where the differences are especially large between the models. However, we find that due to the distance of M87, the effects of gamma-ray absorption are negligible for gamma rays with an energy lower than 10 TeV ($\sim 10^{27}$ Hz). As we predict the emitted flux to peak at $\sim 10^{24-25}$ Hz with a strong flux decrease towards higher energies (see Section \ref{sec:results}), we hence cannot discriminate between any of the three models.



\section{Accretion flow} \label{sec:adaf}

Low-luminosity AGNs like M87, are expected to host accretion flows around their SMBH that are radiatively inefficient. This is characterised by the formation of geometrically thick, optically thin, very hot accretion flows, called Advection-Dominated Accretion Flows (ADAFs, introduced by \citealt{first_adaf_torii, first_adaf} and further developed by e.g. \citealt{Narayan_Yi_original, adafintro}). ADAFs exist only when the accretion rate is sufficiently low ($\dot{M} \lesssim 0.01\dot{M}_\mathrm{Edd}$), and consist of a plasma of thermal electrons and ions, where both components may have different temperatures, $T_e$ and $T_i$ respectively.
In addition to the ADAF, we assume the existence of a truncated standard thin accretion disc (Shakura \& Sunyaev disc, \citealt{ShakuraSunyaev}) extending the outer parts of the ADAF.
Here, we investigate inhowfar an ADAF/disc system can contribute to the X-ray component, while not overshooting the radio-to-optical part of the SED that is considered to be jet dominated.

In the following, we use the quantities $X_n = \frac{X}{10^n}$ and the normalized quantities $r=R/R_\mathrm{S}$, with the Schwarzschild's radius $R_\mathrm{S} = 2 r_g = 2.95 \times 10^5 \, m_\mathrm{BH}$, $m_\mathrm{BH}=M_\mathrm{BH}/M_\odot$ and $\dot{m}=\dot{M}/\dot{M}_\mathrm{Edd} = \eta_\mathrm{eff} \dot{M}c^2/L_\mathrm{Edd}$, where $\eta_\mathrm{eff}$ is the radiation efficiency of the standard thin disk ($\eta_\mathrm{eff} \approx 0.1$) and the Eddington luminosity $L_\mathrm{Edd} \simeq 1.3 \times 10^{47} \, m_{\mathrm{BH},9} \, \mathrm{erg}\, \mathrm{s}^{-1}$ . 
We make use of the one-zone, height-integrated, self-similar solutions of the slim disc equations derived by \cite{Narayan_Yi_original} to describe (see Appendix \ref{appendix:self-solutions}) the hot plasma. 

To obtain the spectrum emitted by an ADAF, the balance between the heating and cooling of the thermalized electrons present in the plasma $ q^{e+} = q^{e-} \label{eq:thermal balance}$, is solved to determine the scaled electron temperature $\theta_e = k_B T_e / m_e c^2$. 
Here $q^{e+}$ is the electrons' heating rate, and $q^{e-}$ is their cooling rate.
The emission mechanisms that we consider in the following are synchrotron radiation, bremsstrahlung and Comptonization of the two previous components. The total cooling rate is the sum of the three individual cooling rates, detailed in Appendices \ref{appendix:brem and synch} and \ref{appendix:compton}. The heating mechanisms and rates are described in Appendix \ref{appendix:heating_rates} and they consist of Coulomb collision between ions and electrons, and viscous energy dissipation.

The plasma is a two-temperature plasma where the ion temperature is related to the electron temperature through $T_i + 1.08T_e \approx 6.66\times 10^{12}\, \beta r^{-1}$ \citep{Narayan_Yi_original}, where $\beta$ is the ratio between the gas $p_g$ and the total pressure $p = \rho c_s^2 = p_m + p_g$ with $p_m = B^2/8\pi$, and $\rho$ is the mass density and $B$ is the isotropically tangled magnetic field.

We obtain the electron temperature by varying $T_e$ using a bisection method to solve the balance equation for each radius.

Furthermore, we take $\dot{m}$ of the form $\dot{m} = \dot{m}_\mathrm{out} \left(r/r_\mathrm{out}\right)^s$, where $r_\mathrm{out}$ is the outer radius of the ADAF and is associated with an accretion rate $\dot{m}_\mathrm{out}$, and $s$ is a mass-loss parameter (introduced by \citep{Blandford_Begelman_massloss}) that is used to include the presence of outflows or winds from the ADAF.



Upon obtaining the electron temperature, the emitted spectrum from the ADAF is computed, integrating over the radius of the ADAF. 
In order to take into account absorption, we follow the method of \cite{ADAF_spectrum_Manmoto}, and derive the flux from synchrotron and bremsstrahlung emission as 
\begin{equation}
    F_{\nu,\mathrm{0}} = \frac{2\pi}{\sqrt{3}}\, B_\nu\left[ 1 - \exp{(-2\sqrt{3}\, \tau_\nu)}\right] \, \, \mathrm{erg} \, \mathrm{cm}^{-2} \, \mathrm{s}^{-1} \, \mathrm{Hz}^{-1} \label{eq:lnu_synch_brem}
\end{equation}
where \begin{eqnarray}
    B_\nu = \frac{2 h \nu^3}{c^2}\frac{1}{e^{\frac{h\nu}{k_\mathrm{B} T_e}} - 1} \nonumber
\end{eqnarray}  is the Planck's function, and $\tau_\nu$ is the optical depth for absorption defined such that $\tau_\nu = (\sqrt{\pi}/2)\kappa_\nu H$, with $\kappa_\nu = (j_{\nu, \mathrm{syn}} + j_{\nu, \mathrm{br}})/(4\pi B_\nu)$ the absorption coefficient. The emissivities $j_{\nu, \mathrm{syn}}$ and $j_{\nu, \mathrm{br}}$ are given in Appendix \ref{appendix:brem and synch}.

Hence the local luminosity from synchrotron and bremsstrahlung at a given radius is given by 
\begin{equation}
    L_{\nu,\mathrm{0}} = 2\pi R^2F_{\nu,\mathrm{0}} 
\end{equation}
Synchrotron radiation and bremsstrahlung further act as a photon field for inverse Compton scattering by the thermal electrons. Following the work of \cite{ADAF_neutrinos_CR}, we compute the number density of photons after the i-th scattering:
\begin{equation}
    N_{\gamma,i}(\epsilon) = \frac{R}{c}\int d\gamma \, \frac{3}{4\gamma^2} \, N_e(\gamma, \theta_e) \, N_{\gamma,i-1}\left(\frac{3\epsilon}{4\gamma^2}\right) \, R_c\left(\frac{3\epsilon}{4\gamma^2}, \gamma\right) \label{eq:compton_flux}
\end{equation}
where $R_c(\epsilon, \gamma)$ is the scattering rate for electrons with Lorentz factor $\gamma$ and photons with dimensionless energy $\epsilon = h\nu/(m_e c^2)$, that we take from \cite{Coppi_Blandford_scattering}. $N_e(\gamma, \theta_e)$ is the Maxwellian distribution of electrons, described in equation \ref{eq:maxwellian}.
The initial condition is given by $N_{\gamma,0}(\epsilon) =  L_{\epsilon,\mathrm{0}}/(h \nu \pi c R^2) $ with $L_{\epsilon,\mathrm{0}} = (m_e c^2/h) L_{\nu,\mathrm{0}} $.

The self-similar solutions give a good estimate for the ADAF emission for sufficiently large radii ($r \gg r_\mathrm{sonic}$, where $r_\mathrm{sonic}$ is the sonic radius; \citealt{Sonic_radius_critical}), however the inner part of the ADAF ($r \sim 2.5 - 4$) is thought to be at the origin of the ring observed by the EHT collaboration \citep{EHT2019_BZ_BP} at 230 GHz. We cannot use the self-similar solutions to account for this inner part emission, but, considering the ADAF framework, we expect that synchrotron radiation is the dominant process in this region. The synchrotron radiation is self-absorbed until the peak frequency corresponding to the emission radius (here 230 GHz at $r \sim 2.5 - 4$ corresponds to $R \sim 5 - 8 \, r_g$), hence we add a power-law component $F_\nu \propto \nu^{5/2}$ that we scale to the observed flux at 230 GHz, to the existing ADAF spectrum (coming from regions $r \ge 5$).
The maximum radius $r_\mathrm{max}$ is poorly constrained. As there is no evidence for the presence of a truncated thin disc in the infrared data, we set $r_\mathrm{max} = 2\times 10^5$ so that any contribution from an outer disc truncated at this radius would be negligible (the computation of the outer disc spectrum is performed in Appendix \ref{appendix disc}). This value is consistent with the Bondi radius derived by \cite{bondiradius}. For the remaining parameters, we explore a broad range of values, as summarized in Table \ref{table:parameters_ADAF}.

\begin{table}
    \centering
    \tablenum{1}
    \label{table:parameters_ADAF}
     \caption{Summary of the allowed ranges for the parameters $\alpha$, $\beta$, $\dot{m}_\mathrm{out}$, s, $\delta_e$, as well as the best values chosen to represent the ADAF model for M87. Here $\alpha$ is the viscosity parameter introduced by \cite{ShakuraSunyaev}, $\beta$ is the ratio between the gas and the total pressure, $\dot{m}_\mathrm{out}$ is the accretion rate at the outermost part of the ADAF, s is the mass-loss parameter characterizing the evolution of the accretion rate over the volume, and $\delta_e$ is the fraction of viscous energy directly transmitted to the plasma electrons.}
    \begin{tabular}{ccccc}
    \hline
    \hline
         parameter & minimum value & maximum value & best choice & reference work \\
         \hline
         $\alpha$ & 0.01 & 1 & 0.1  & 1, 2 \\ 
         $\beta$ & 0.5 & $<1$ & 0.9 & 3, 4   \\
         $\dot{m}_\mathrm{out} $ & $1\times 10^{-4}$ & $2.93\times 10^{-3}$  \tablenotemark{a} & $1.6 \times 10^{-3}$  & 5, 6 \\ 
         $s$ & 0 & 1 &  0.39 & 6, 7 \\ 
         $\delta_e$ & $10^{-4}$ & $10^{-1}$ & $5 \times 10^{-3}$ & 5, 8\\
    \hline
    \end{tabular}
    \tablenotetext{a}{ The upper limit is the Bondi accretion rate, calculated with the mass estimate of \cite{2017_Mass_EHT}}
    \tablerefs{(1) \citealt{alpha_visco_mad}, (2) \citealt{alpha_visco_obs}, (3) \citealt{beta_values1}, (4) \citealt{beta_values2}, (5) \citealt{DiMatteoM87}, (6) \citealt{LLAGNmodels}, (7) \citealt{Blandford_Begelman_massloss}, (8) \citealt{Mahadevan1997}}
\end{table}

\begin{figure}[ht]
     \centering
     \includegraphics[width = 0.8\textwidth]{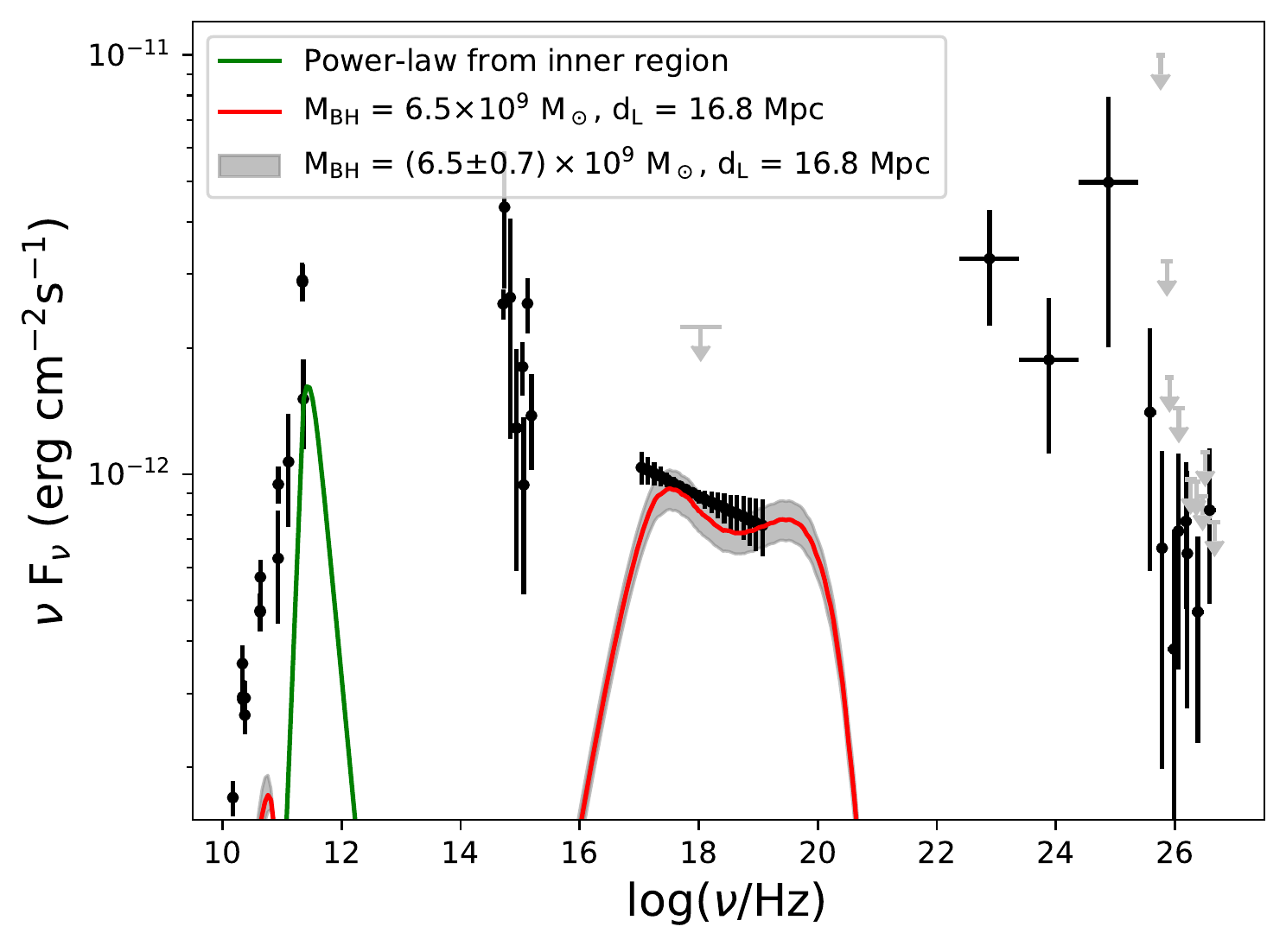}
     \caption{SED from the ADAF component. The model with the best-fitting parameters Table~(\ref{table:parameters_ADAF}) with M$_\mathrm{BH} = 6.5 \times 10^9 M_\odot$ is given by the red line. The grey area represents the variation of this model for the range of mass M$_\mathrm{BH} = (6.5\pm 0.7) \times 10^9 M_\odot$}
     \label{fig:adaf}
 \end{figure}

 For this work, we wish to probe whether an ADAF component could explain the X-ray data, without overestimating the radio to optical observations. In Figure \ref{fig:adaf}, we present the spectrum obtained with the parameter values that represent the data best. With the accretion rate dependency on the radius, its value in the innermost regions $r\sim 2.5$ is set to the value inferred from the black hole ring observations \citep{EHT2019_BZ_BP}, where an accretion rate in the inner region of $\dot{m}\sim 2\times 10^{-5}$ was estimated. The values of $\beta$ and the electron density in the black hole vicinity are compatible with values derived for MAD (Magnetically Arrested Disk; see e.g. \citealt{firstMAD, MAD}) simulations \citep{MagFieldEHT}. The ADAF component alone is not entirely consistent with the X-ray data, however its contribution is added to the jet emission to produce the overall SED (see Section \ref{sec:results}).

\section{Results} \label{sec:results}

With the methods described above we probe whether the total joint model (jet component added to the ADAF component) can explain the global SED. We start by setting the fixed parameters of the jet. Since the synchrotron self-absorption frequency is a critical feature of the observed spectrum (see Section \ref{sec:jet section}), we fix the size of the emission region in order to maximise the self-absorption frequency value, while being consistent with the measured \cite{EHT2017A} flux value, namely we set $r_\mathrm{em}' = 5\times 10^{15}$ cm $\approx 5 \, r_g$. This corresponds to the radius of the sphere in the moving blob scenario, while for the continuous jet this gives the transverse radius of the cylinder.

For this region, we explore a parameter space starting with varying the magnetic field strength between 10~G and 50~G. For each magnetic field we adjust the electron maximum energy and spectral index in order to reproduce the observed cutoff in the optical band, while complementing the ADAF contribution around $10^{16}$ Hz.
Once we find the combination between the jet magnetic field strength, the size of the emission region and the injection rate of electrons inside the jet region we determine the Doppler factor $\delta_j \approx 2.3$. 
This corresponds to a velocity $\beta c = 0.73c$ with an inclination of the jet $\theta = 17 \degr$. With this value of the Doppler factor and the size of the emission region we choose, the length of the cylinder, using the geometry described in \ref{sec:jet section}, is $l' \approx 10^{16} \, \mathrm{cm}$. 
For the injected proton population, we explore cutoff energies between $10^9$ and $10^{10}$ GeV, and spectral indices between 1.7 and 2.0. There are less observational constraints on the proton population than for the electrons. We check the ratio of the maximum proton-to-electron energy (see Section \ref{sec:jet section}) and consider models for which the total energy density in particles is lower than or equal to the magnetic energy density.

\begin{figure}[ht]
     \centering
     \includegraphics[width = 0.9\textwidth]{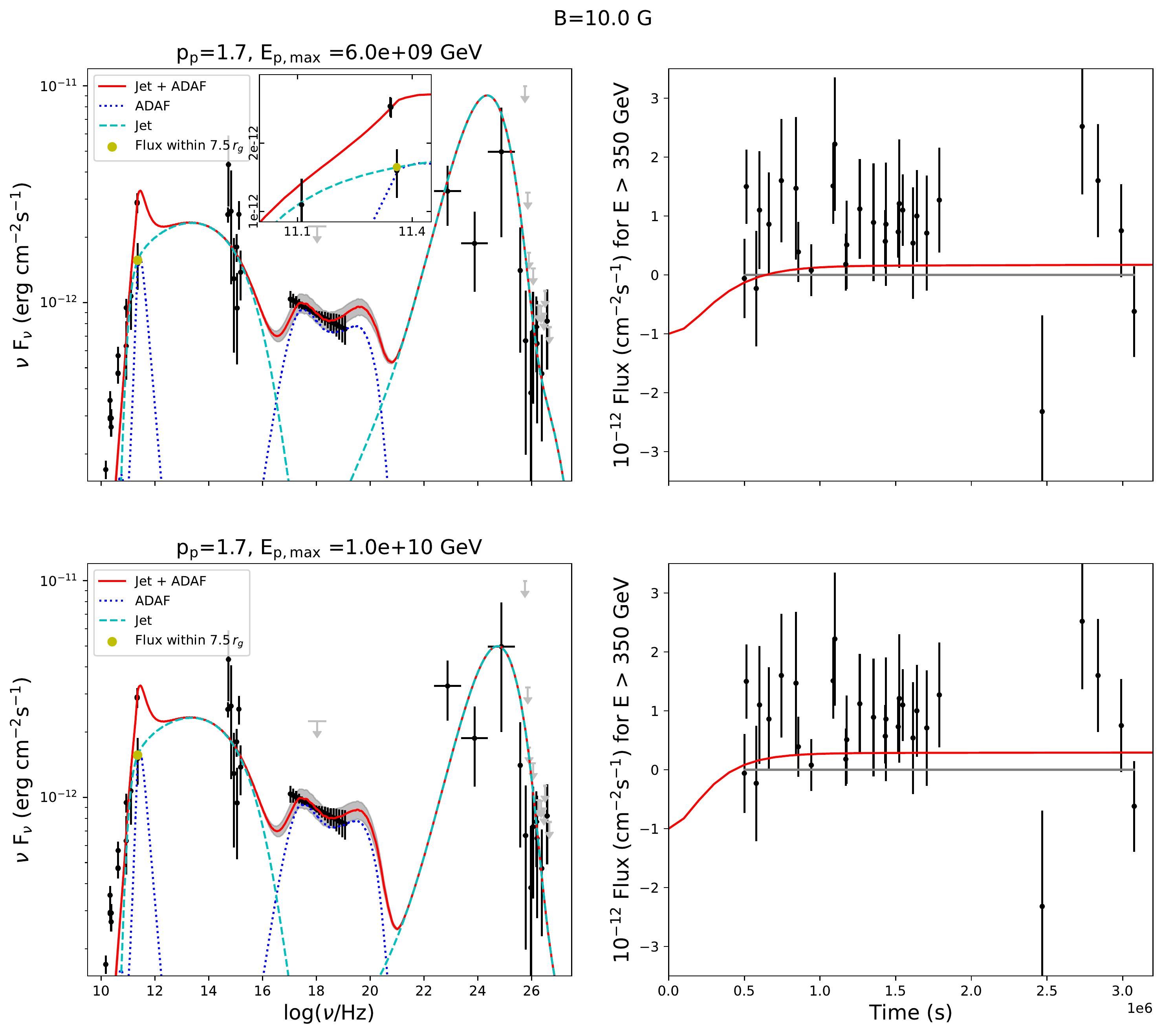}
     \caption{Left panels: Global core emission (red line), composed of the jet emission (cyan dashed curve) and the ADAF emission (blue dotted curve), for a proton spectral index $p_\mathrm{p} = 1.7$ and maximum proton energy $E_\mathrm{p,max} = 6\times 10^9$ GeV (top panel) and  $E_\mathrm{p,max} = 10^{10}$ GeV (bottom panel), for a magnetic field of 10 G. Right panels: Corresponding integrated light curve for energies $E > 350$ GeV (red lines). The grey straight line represents the estimated level of flux with no variability in \cite{EHTpaper}.}
     \label{fig:flux_B10G_p170}
 \end{figure}

As mentioned in Section \ref{sec:jet section}, the accretion flow could serve as an external target photon field for the jet's interactions. To assess if the ADAF would make a relevant target field, we compare the energy density of the internal (jet) and external (flow) photon fields in the jet's frame. To do so we transform the accretion flow radiation field into the jet's frame, assuming for simplicity that the the flow is seen as a point source behind the jet. This is a rough approximation, however we only want to estimate the dominant field here.
\begin{figure}[ht]
    \centering
    \includegraphics[width = 0.7\textwidth]{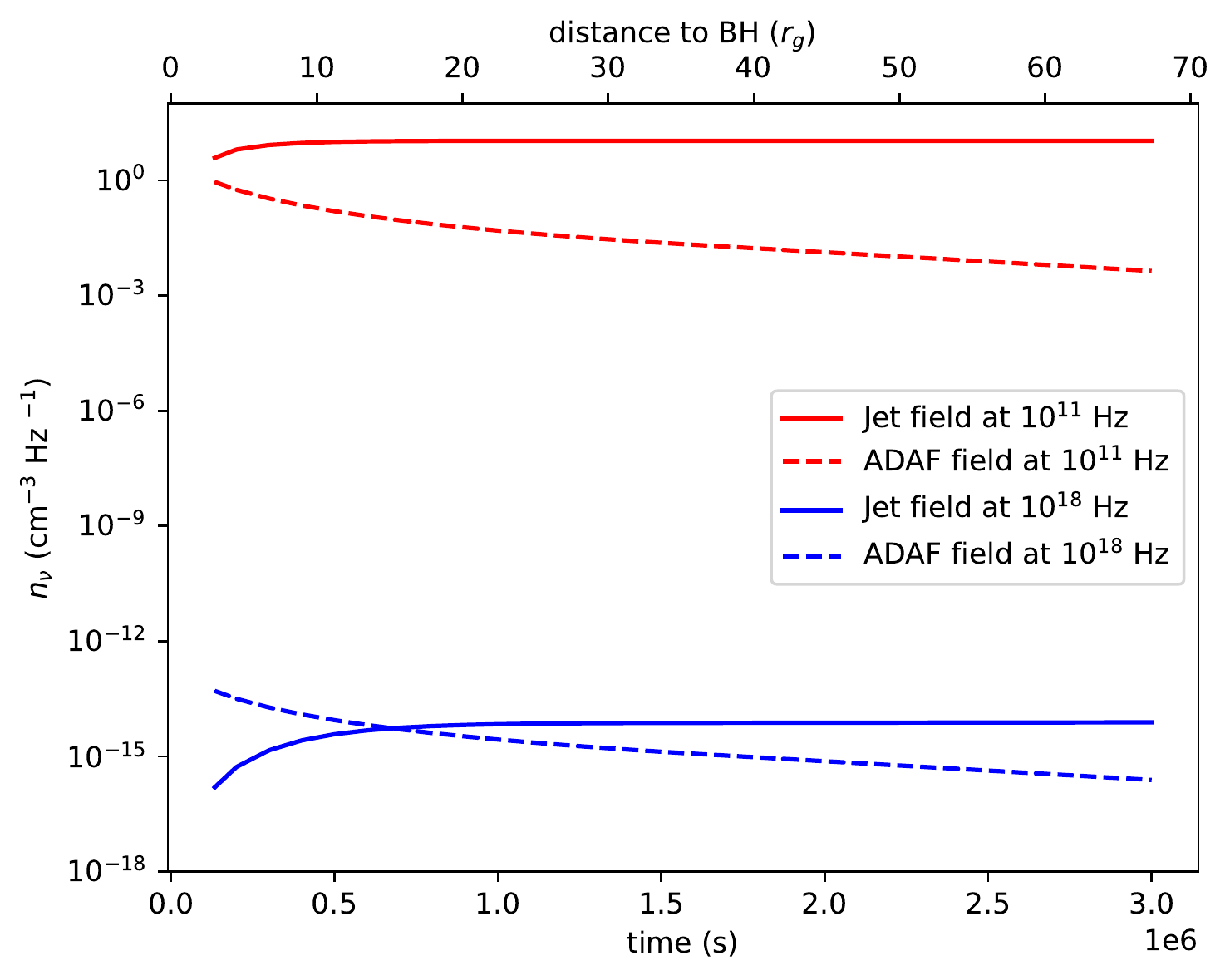}
    \caption{Energy density of the targets fields in the jet's frame at $10^{11}$ Hz (red) and $10^{18}$ Hz (blue). The solid and dashed lines represent the contribution from the jet and the accretion flow, respectively.}
    \label{fig:target_fields_comparison}
\end{figure}
 In Figure \ref{fig:target_fields_comparison}, we compare the photon spectral number density of the photon fields (for the jet model for which $B = 10\,\mathrm{G}$, $p_{p} = 1.7$, $E'_{\mathrm{max},p}=6\times 10^9\,\mathrm{GeV}$, corresponding to the top panel of Figure \ref{fig:flux_B10G_p170}, and the ADAF shown in Figure \ref{fig:adaf}) at two frequencies $10^{11}$ Hz and $10^{18}$ Hz, at which we expect the accretion flow to contribute (see Section \ref{sec:adaf}). Obviously, the internal radio photon field is dominating the external radio photon field. Even at X-ray energies, after only a short time (10 days, over the two months of simulated observation) the internal target field contribution is larger than the external one. Therefore we do not consider the accretion flow as an external target photon field for the jet particles.

The SED is obtained by averaging the light curves over a time corresponding to the observation campaign time, i.e. 2 months.
The goodness of the fits (for both the light curve and the SED) is estimated by computing the p-value of the $\chi^2$-test for each model. We keep models that have a p-value $p_{\chi^2} > 0.01$.

 \begin{figure}[ht]
     \centering
     \includegraphics[width = 0.9\textwidth]{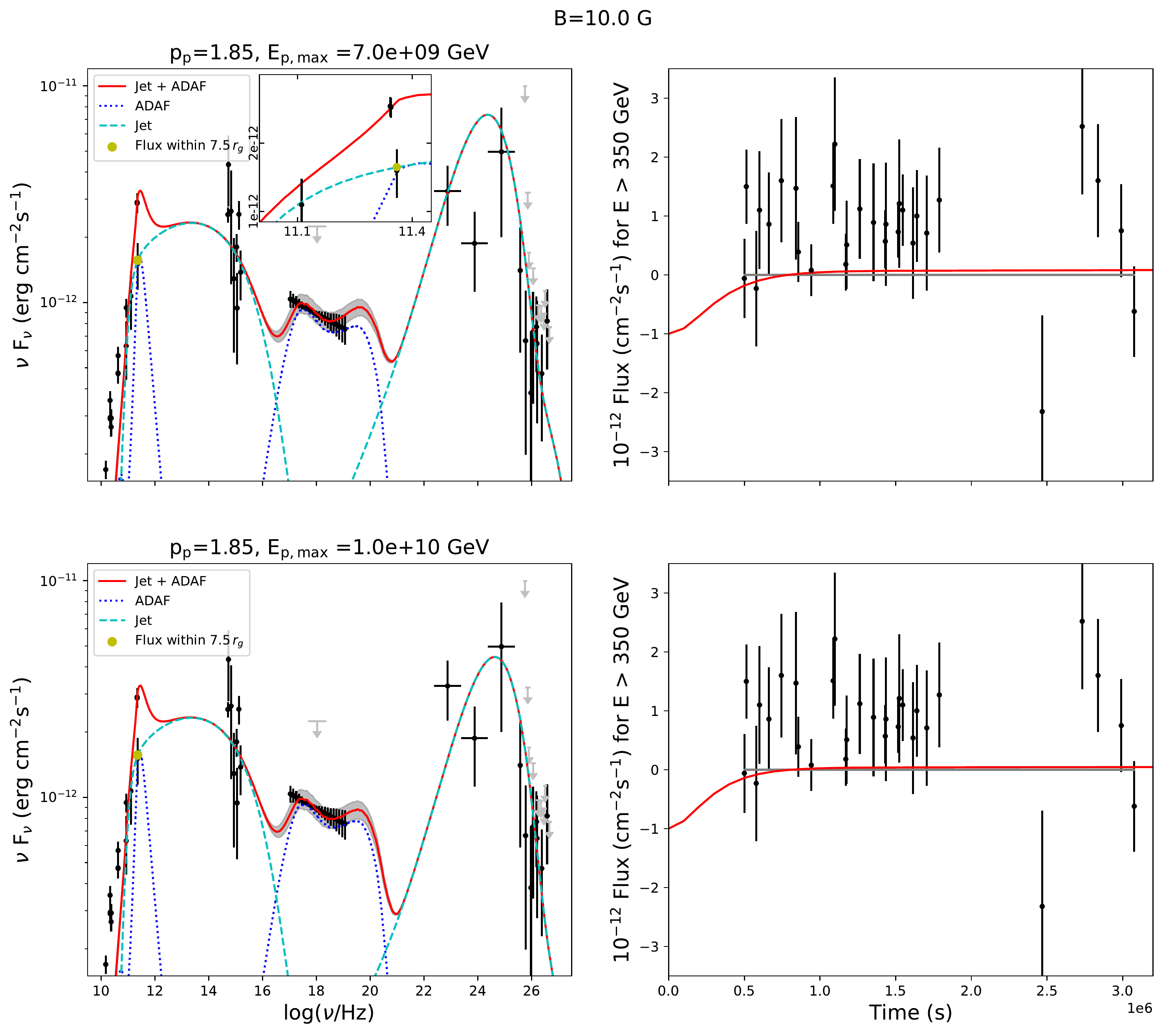}
     \caption{Same as figure \ref{fig:flux_B10G_p170}, but for a proton spectral index $p_\mathrm{p} = 1.85$ and maximum proton energy $E_\mathrm{p,max} = 7\times 10^9$ GeV (top panel) and  $E_\mathrm{p,max} = 10^{10}$ GeV (bottom panel).}
     \label{fig:flux_B10G_p185}
 \end{figure}
 
 \begin{figure}[ht]
     \centering
     \includegraphics[width = 0.9\textwidth]{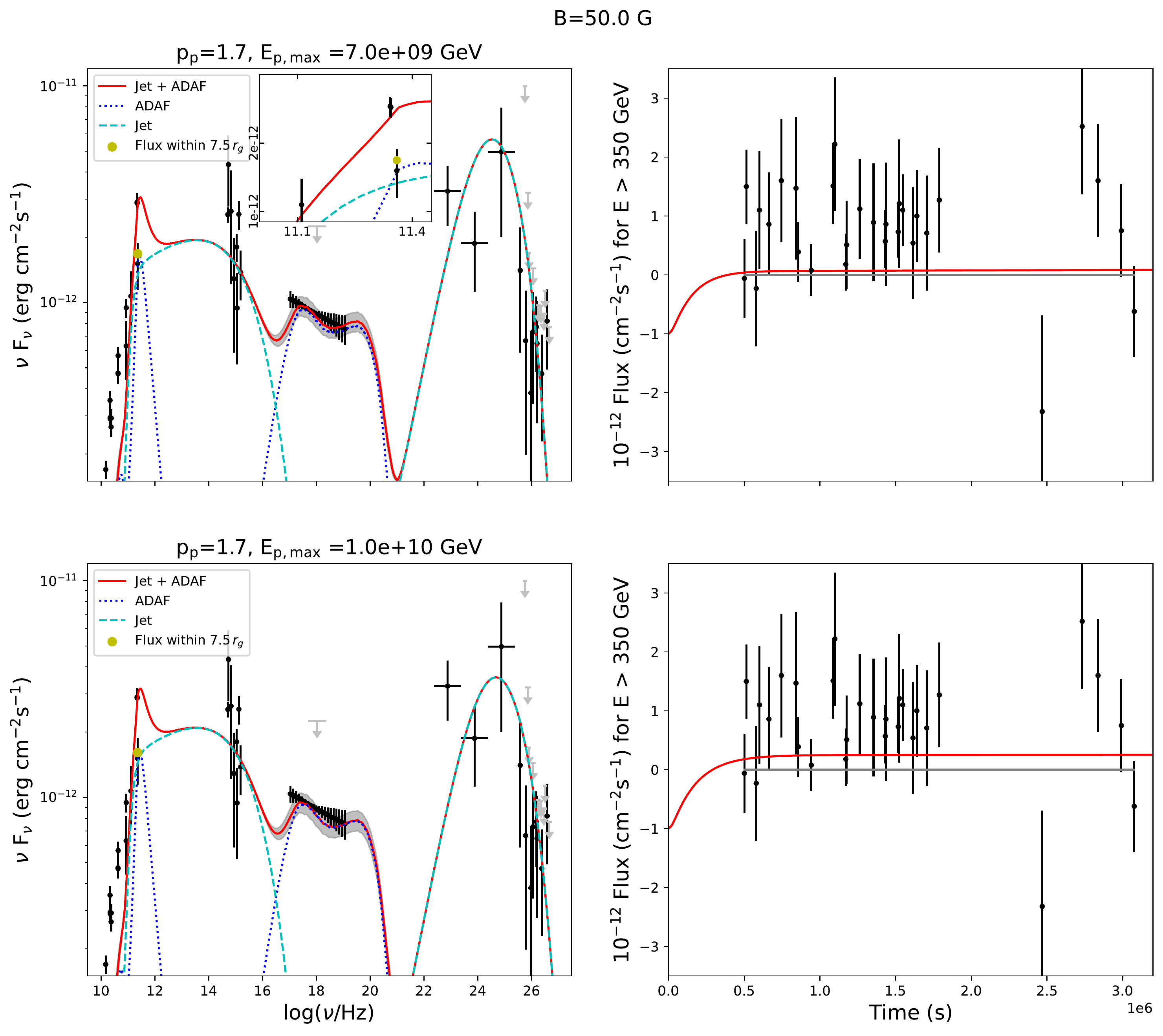}
     \caption{Left panels: Global core emission (red line), composed of the jet emission (cyan dashed curve) and the ADAF emission (blue dotted curve), for a proton spectral index $p_\mathrm{p} = 1.7$ and maximum proton energy $E_\mathrm{p,max} = 7\times 10^9$ GeV (top panel) and  $E_\mathrm{p,max} = 10^{10}$ GeV (bottom panel), for a magnetic field of 50 G. Right panels: Corresponding integrated light curve for energies $E > 350$ GeV (red lines). The grey straight line represents the estimated level of flux with no variability in \cite{EHTpaper}.}
     \label{fig:flux_B50G_p170}
 \end{figure}
 
 \begin{figure}
     \centering
     \includegraphics[width = 0.9\textwidth]{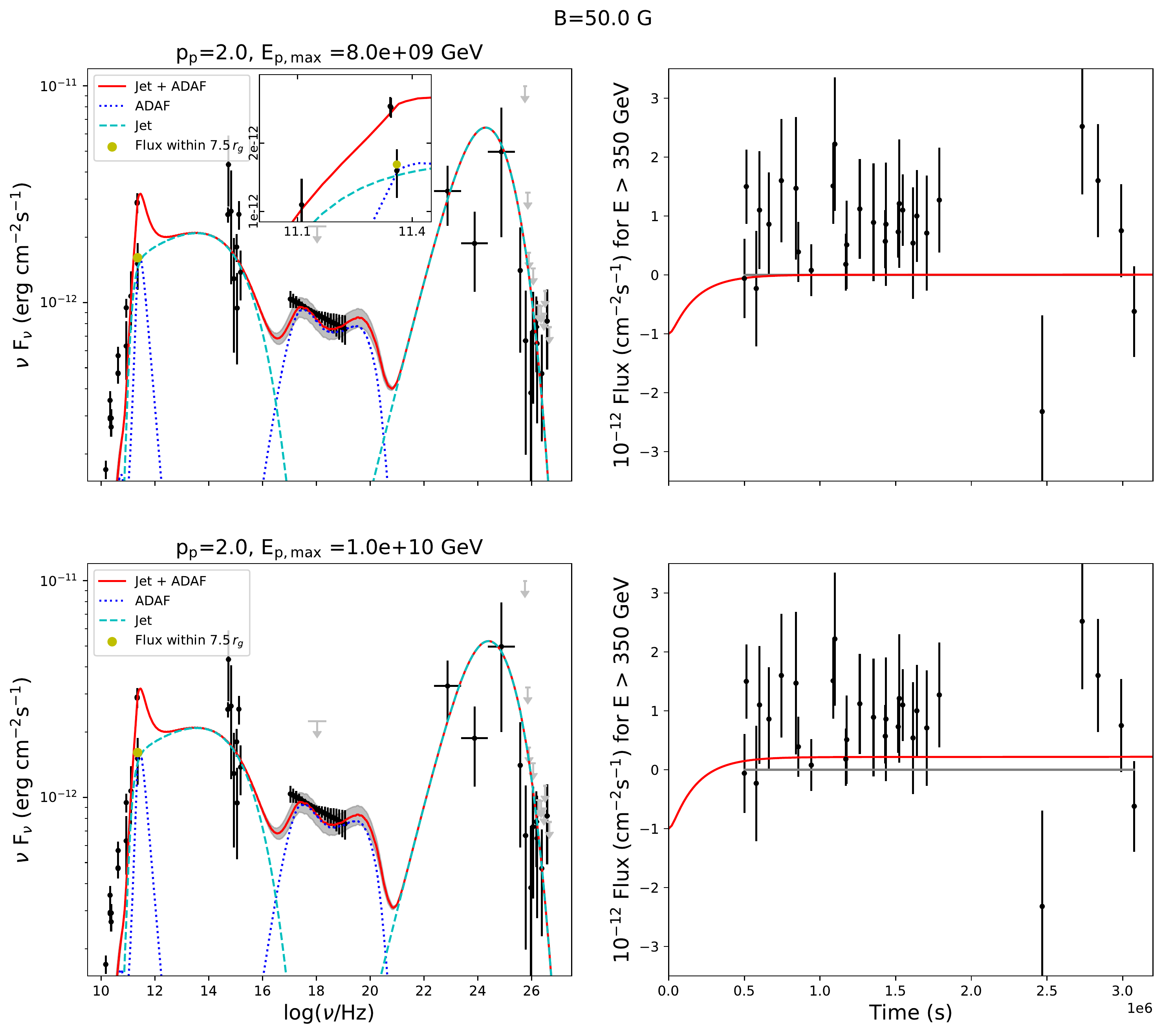}
     \caption{Same as figure \ref{fig:flux_B50G_p170}, but for a proton spectral index $p_\mathrm{p} = 2$ and maximum proton energy $E_\mathrm{p,max} = 8\times 10^9$ GeV (top panel) and  $E_\mathrm{p,max} = 10^{10}$ GeV (bottom panel).}
     \label{fig:flux_B50G_p200}
 \end{figure}

In Figures \ref{fig:flux_B10G_p170} and \ref{fig:flux_B10G_p185} we present four models for which $B = 10$~G. The models have the lowest (highest) maximum proton energy and lowest (highest) proton index possible given the observations and the constraints listed in Section \ref{sec:jet section}. With these models we obtain a jet power of $P_\mathrm{j} = 2-4 \times 10^{43} \, \mathrm{erg}\, \mathrm{s}^{-1}$ and ratios of magnetic-to-particle energy density of $U_\mathrm{part}/U_B = 0.6-1.3$.
In Figures \ref{fig:flux_B50G_p170} and \ref{fig:flux_B50G_p200} we did the same exploration, and present four models for which $B = 50$~G. We find that with such a high value for the magnetic field strength, it is harder to fit the data, and one has to consider lower proton densities and higher maximum proton energies. For a proton injection spectrum of $p=2$, we find a good fit only for maximum proton energies $E_\mathrm{p,max} \ge 8\times 10^{9} $~GeV (see top panel in Figure \ref{fig:flux_B50G_p200}). With these models we obtain a jet power of $P_\mathrm{j} \approx 3 \times 10^{44} \, \mathrm{erg}\, \mathrm{s}^{-1}$ and ratios of magnetic energy density to particle energy density of $U_\mathrm{part}/U_B \approx 10^{-2}$

For both $B=10$~G and $B=50$~G, it is easier to obtain a light-curve above 350 GeV complying with the observations with a higher value of the maximum proton energy, but since proton synchrotron radiation represents the main contribution to the high-energy spectral bump, the higher the maximum proton energy, the higher the frequency the emission will peak at, and the SED fits get poorer.

\begin{figure}[ht]
    \centering
    \includegraphics[width = 0.8\textwidth]{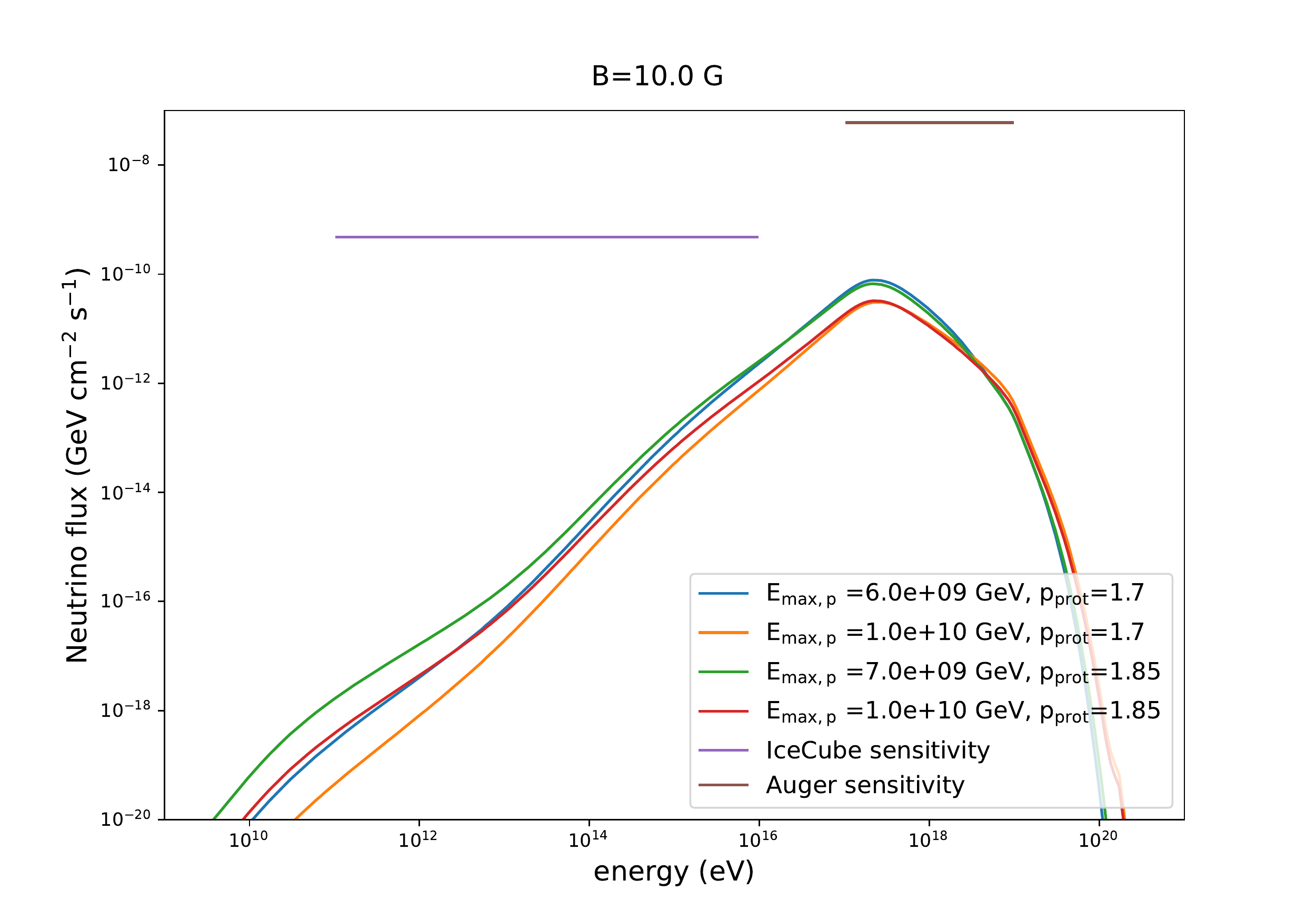}
    \caption{Predicted electron and muon neutrino spectra, for the models with $B=10$ G.}
    \label{fig:neutrinos}
\end{figure}

The neutrino spectra (single-flavor flux) produced by the source from the models with $B = 10$~G are presented in Figure \ref{fig:neutrinos}. The predicted flux is low, because the main gamma-ray emission contribution is due to proton synchrotron radiation, which does not produce neutrinos. We compare this value to the sensitivity to a point-like source of high-energy neutrinos with a neutrino flux $\propto E^{-2}$, of the Pierre Auger Observatory \citep{PierreAuger} and the IceCube observatory \citep{IceCube} at M87's declination. 

\section{Conclusions} \label{sec:conclusion}

We have applied a lepto-hadronic, time-dependent jet model, complemented with an advection dominated accretion flow to M87's nuclear emission in a low flux state. We found a range of parameter values that allow to reproduce the multi-wavelength data taken in 2017 during the \cite{EHTpaper} observation campaign. We investigated two types of jet configuration, namely the moving blob and the continuous jet scenario. For a given geometry, we were able to find identical results for both geometries. We focused on an jet emission region of the size similar to the EHT angular resolution at M87's distance, namely $5 r_\mathrm{g}$. Within this region we estimated a magnetic field strength in the range 5-60 G. The level of flux around the synchrotron self-absorption frequency ($86 < \nu_\mathrm{SSA} < 230$ GHz) constrains further the injection parameters of the relativistic electrons in the jet. For a range $10 \, \mathrm{G} \le B \le 50 \, \mathrm{G}$ we found that the electrons spectral index is limited to $p_e \approx 1.80 - 1.85$, in order to reproduce the radio-to-optical part of the SED. For the same reason, the maximum energy for the electron distribution is found to be $E_{\mathrm{max},e} \lesssim 5$ GeV. Concerning the high energy emission, we have found parameter values that fit the data for the whole range of magnetic field strengths considered. However, it it worth pointing out that the proton maximum energy and spectral index ranges are dependent on the value for the magnetic field strength. For $B = 10$ G we found that when $p_p \simeq 1.7$ (minimum proton spectral index) the proton maximum energy is in the range $6\times 10^9 \, \mathrm{GeV} \lesssim E_{\mathrm{max}, p} \lesssim 1\times 10^{10} \, \mathrm{GeV}$ while for $p_p \simeq 1.85$ (maximum proton spectral index for these parameter values) it is in the range $7\times 10^9 \, \mathrm{GeV} \lesssim E_{\mathrm{max}, p} \lesssim 1\times 10^{10} \, \mathrm{GeV}$. For $B = 50$ G when $p_p \simeq 1.7$ (minimum proton spectral index) the proton maximum energy is in the range $7\times 10^9 \, \mathrm{GeV} \lesssim E_{\mathrm{max}, p} \lesssim 1\times 10^{10} \, \mathrm{GeV}$ while for $ p_p \simeq 2.00 $ (maximum proton spectral index for these parameter values) it is in the range  $8\times 10^9 \, \mathrm{GeV} \lesssim E_{\mathrm{max}, p} \lesssim 1\times 10^{10} \, \mathrm{GeV}$. This required increase in the maximum proton energy makes it harder to fit the gamma-ray part of the SED above $10^{25}$ Hz.
Combining the jet's emission with the ADAF allows us to reproduce at the same time the apparent cut-off in the optical band, and the power-law-like flux component at X-rays energies.
Unlike previous works (e.g. \citealt{Feng_accretion_jet_model, LLAGNmodels}), we found a configuration where both the jet and the ADAF have a distinct contribution in the SED. Beyond the scope of this paper, as mentioned in the introduction, would be the estimation of the contribution to the cosmic-ray flux from M87. Indeed, with protons accelerated up to $10^{10}\, \mathrm{GeV}$ and a jet power of $10^{43-44}$ erg s$^{-1}$, M87 could contribute to the detected high-energy cosmic-ray flux on Earth (\cite{M87UHECR_TeV}, see \citealt{uhecr_power_requirement} about the power requirements of cosmic-ray sources).

In this work we have considered only one-zone models for the jet emission. In the framework of structured jet models, multi-zone scenarios have been invoked to explain M87's SED (e.g. \citealt{Two-flow_jets, structuredjet}). In particular, \cite{SpineSheath_jets} developed a leptonic scenario in which a fast inner jet is embedded in a slower outer sheath. Here, the beaming pattern related to the boosting of one layer into the other could explain the high-energy part of the SED. They applied it to M87's SED \citep{M87_spine_sheath}. A transverse structure in jets is further supported by observations of limb-brightening (for M87 see \citealt{M87_limb}, for radio-galaxies and blazars see \citealt{Mrk_structure, RG_structure}). However, with twice as many parameters as for one-zone jet models, such as the one we considered, it is difficult to constrain two-zone scenarios to date.
\\

\begin{acknowledgments}
MB has for this project received funding from the European Union’s Horizon 2020 research and innovation program under the Marie Sklodowska-Curie grant agreement No 847476. The views and opinions expressed herein do not necessarily reflect those of the European Commission. MB wishes to thank Paolo Da Vela for the fruitful discussions and insightful comments on this paper. 

LM acknowledges support from the DFG within the Collaborative Research Center SFB1491 "Cosmic Interacting Matters - From Source to Signal".

This research was funded in part by the Austrian Science Fund (FWF)
(grant number I 4144-N27). For the purpose of open access, the author has applied a CC BY public copyright licence to any Author Accepted Manuscript version arising from this submission.

We would like to thank the anonymous referee for comments and suggestions that helped improve this paper.

\software{This work benefited from the following software: NumPy \citep{numpy}, Matplotlib \citep{matplotlib}, pandas \citep{pandas, panda_software}, jupyter notebooks \citep{ipython}.}

\end{acknowledgments}


\appendix

\section{Jet's magnetic field: b(p) coefficient} \label{b(p)}

\cite{Kino_SSA} derived a relation between the jet's magnetic field and the observable quantities:

\begin{eqnarray*}B = b(p)\left(\frac{\nu_\mathrm{ssa,obs}}{1\mathrm{GHz}}\right)^5\left(\frac{\theta_\mathrm{obs}}{1\mathrm{mas}}\right)^4\left(\frac{S_{\nu_\mathrm{ssa},\mathrm{obs}}}{1\mathrm{Jy}}\right)^{-2}\left(\frac{\delta}{1+z}\right) \, \mathrm{G}.
\end{eqnarray*}
Here $b(p)$ is defined as
$ b(p) = 5.52\times 10^{57}\, \Big[\Big(3 X_2 \, c_2(p)\Big)/\Big(2\pi X_1 \, c_1(p)\Big)\Big]^2 $, where 
\begin{eqnarray*}
    X_1 = \frac{\sqrt{3}e^3}{8\pi m_e}\left(\frac{3e}{2\pi m_e^3 c^5}\right)^{p/2}
\end{eqnarray*}

\begin{eqnarray*}
    c_1(p)=\Gamma\left(\frac{3p + 2}{12}\right)\Gamma\left(\frac{3p + 22}{12}\right)
\end{eqnarray*}

\begin{eqnarray*}
    X_2 = \frac{\sqrt{3}e^3}{8\sqrt{\pi} m_e c^2}\left(\frac{3e}{2\pi m_e^3 c^5}\right)^{(p-1)/2}
\end{eqnarray*}

\begin{eqnarray*}
    c_2(p)=\Gamma\left(\frac{3p + 19}{12}\right)\Gamma\left(\frac{3p -1}{12}\right)\Gamma\left(\frac{p +5}{4}\right)/\Gamma\left(\frac{p + 7}{4}\right)/(p+1)
\end{eqnarray*}

\section{ADAF self-similar solutions} \label{appendix:self-solutions}

The one-zone, height-integrated, self-similar solutions of the slim disc equations were derived by \cite{Narayan_Yi_original} to describe the hot plasma. The solutions and their expression using the relevant scaled quantities are given by:

\begin{eqnarray}
    v_R &\approx& \frac{\alpha}{2}v_K \approx 1.06\times 10^9 \, \alpha_{-1} \, r^{-1/2} \,\, \mathrm{cm} \, \mathrm{s}^{-1} \nonumber \\
    c_s &\approx& \frac{1}{2} v_K \approx 1.06\times 10^{10} \, r^{-1/2} \,\, \mathrm{cm} \, \mathrm{s}^{-1} \nonumber\\
    H &\approx& \frac{1}{2}R \approx 1.48 \times 10^{14} \, m_{\mathrm{BH},9} \, r \,\, \mathrm{cm}  \\
    \rho  = n_p\,m_p &\approx& \frac{\dot{M}}{4\pi R H v_R } \approx 2.66\times 10^{-15} \, m_{\mathrm{BH},9}^{-1} \, \dot{m}_{-3} \,  \alpha_{-1}^{-1} \, r^{-3/2} \,\, \mathrm{g} \, \mathrm{cm}^{-3} \nonumber\\
    B &\approx& \sqrt{8\pi \rho c_s^2 (1 - \beta)} \approx 2.75 \times 10^3 \, m_{\mathrm{BH},9}^{-1/2} \, \dot{m}_{-3}^{1/2} \, \alpha_{-1}^{-1/2} \, (1-\beta)^{1/2} \, r^{-5/4} \,\, \mathrm{G} \nonumber\\
    \tau_\mathrm{T} &=& n_p \sigma_\mathrm{T} R \approx 0.313 \, \dot{m}_{-3} \, \alpha_{-1}^{-1} \, r^{-1/2} \nonumber
\end{eqnarray}
where $v_K = \sqrt{G M_\mathrm{BH}/R}$ is the Keplerian velocity, $v_R$ is the radial velocity, and $c_s$ is the isothermal sound speed. Here, $\alpha$ is the viscosity parameter introduced by \cite{ShakuraSunyaev}, $\beta$  the ratio between the gas $p_g$ and the total pressure $p = \rho c_s^2 = p_m + p_g$ with $p_m = B^2/8\pi$, where $B$ is the isotropically tangled magnetic field. The Thomson optical depth is denoted by $\tau_\mathrm{T}$.

\section{ADAF cooling mechanisms} \label{appendix:details_synch_brem}

\subsection{Synchrotron radiation and bremsstrahlung} \label{appendix:brem and synch}

We assume that the plasma electrons follow a relativistic Maxwellian distribution
\begin{equation}
    N_e(\gamma_e, \theta_e) = n_e \frac{\gamma_e^2 \beta_e \exp{(-\gamma_e/\theta_e)}}{\theta_e \, K_2(1/\theta_e)} \label{eq:maxwellian},
\end{equation} where $n_e \approx n_p$ is the electrons number density, $\beta_e$ and $\gamma_e$ are the relative velocity and the Lorentz factor of the thermal electrons respectively and $K_n(x)$ is the n-th order modified Bessel function.

For synchrotron radiation from thermal electrons and bremsstrahlung, we use the fitting formula derived by \cite{Narayan_Yi_original}. The synchrotron emissivity is given by 
\begin{equation}
    j_{\nu, \mathrm{syn}} =  4.43\times 10^{-30}\frac{4\pi \, n_e \, \nu}{K_2(1/\theta_e)} \, I'\left(\frac{4\pi \, m_e \, c \, \nu}{3 \, e \, B \, \theta_e^2}\right) \, \, \mathrm{erg} \, \mathrm{cm}^{-3} \, \mathrm{s}^{-1} \, \mathrm{Hz}^{-1}\label{eq:syn_emissivity}
\end{equation}
where $I'(x)$ is defined in \cite{Narayan_Yi_original}:
\begin{eqnarray*}
    I'(x) = \frac{4.0505}{x^{1/6}}\left( 1 + \frac{0.4}{x^{1/4}} + \frac{0.5316}{x^{1/2}} \right)\exp{(-1.8899\,x^{1/3})}
\end{eqnarray*}

The Bremsstrahlung cooling rate is given by the sum of the rates from electron-electron and ion-electron interactions \citep{rates_hotplasma, Svensson_brems}: 
\begin{equation}
    q_\mathrm{br} = q_\mathrm{ee} + q_\mathrm{ei}
\end{equation}

The ion-electron and  electron-electron bremsstrahlung cooling rates are respectively given by \citep{Svensson_brems, rates_hotplasma}:
\begin{eqnarray*}
    q_\mathrm{ei} & = & 1.48 \times 10^{-22} \, n_e^2 \, F_\mathrm{ei}(\theta_e) \, \, \mathrm{erg} \, \mathrm{cm}^{-3} \, \mathrm{s}^{-1}\\
    q_\mathrm{ee} & = & \left\{
    \begin{array}{cc}
         & 2.56 \times 10^{-22} \, n_e^2 \, \theta_e^{3/2} \, (1 + 1.1\theta_e + \theta_e^2 - 1.25\theta_e^{5/2}) \quad \mathrm{if} \, \, \theta_e < 1\\
         & 3.40 \times 10^{-22} \, n_e^2 \, \theta_e \, \left[ \ln{(1.123\theta_e)} + 1.28 \right]  \quad \mathrm{if} \, \, \theta_e > 1
    \end{array}
    \right. \, \, \mathrm{erg} \, \mathrm{cm}^{-3} \, \mathrm{s}^{-1} 
\end{eqnarray*}
where 
\begin{eqnarray*}
    F_\mathrm{ei}(\theta_e) = \left\{
    \begin{array}{cc}
         & 4\left( \frac{2\theta_e}{\pi^3} \right)^{0.5} \, (1 + 1.1781\theta_e^{1.34}) \quad \mathrm{if} \, \, \theta_e < 1\\
         & \frac{9\theta_e}{2\pi} \, \left[ \ln{(1.123\theta_e + 0.48)} + 1.5 \right] \quad \mathrm{if} \, \, \theta_e > 1
    \end{array}
    \right.
\end{eqnarray*}

Assuming a Gaunt factor equal to unity, we approximate the emissivity due to bremsstrahlung to be 
\begin{equation}
    j_{\nu, \mathrm{br}} \approx q_\mathrm{br}\frac{h}{k_\mathrm{B} T_e}\exp{(-\frac{h\nu}{k_\mathrm{B}T_e})} \,. 
\end{equation}

\subsection{Compton cooling} \label{appendix:compton}

In order to take into account the inverse Compton scattering in the cooling mechanism we use the formulation derived by \cite{Esin_compton_enhancement} to compute the cooling rate. 
Assuming the comptonization is enhancing the initial energy of the seed photons, we can define the energy enhancement factor $\eta(\nu)$ such that

\begin{equation}
    \eta = \exp{[s(A-1)]}[1 - P(j_m+1, As)] + \eta_\mathrm{max}P(j_m+1, s)
\end{equation}

where $P(a,x)$ is the regularized lower incomplete gamma function and
\begin{eqnarray*}
    A = 1 + 4\theta_e + 16\theta_e^2, \quad s = \tau_\mathrm{T} + \tau_\mathrm{T}^2 \\
    \eta_\mathrm{max} = \frac{3k_\mathrm{B} T_e}{h\nu}, \quad j_m = \frac{\ln{\eta_\mathrm{max}}}{\ln{A}} \quad .
\end{eqnarray*}

Finally the total cooling rate of the electrons is given by
 \begin{equation}
     q^{e-} = \frac{1}{H} \int \mathrm{d}\nu \eta(\nu) F_{\nu,\mathrm{0}} \label{eq:Qminus}
 \end{equation}
 where $F_{\nu,\mathrm{0}}$ is described in the main text, equation \ref{eq:lnu_synch_brem}.

\section{ADAF heating rates}  \label{appendix:heating_rates}

The electrons are heated in two different ways in the plasma, namely they can be directly heated by a fraction $\delta_e$ of the viscous dissipated energy, and they can also be heated through Coulomb collisions with the ions.

The viscous energy dissipation rate per unit volume $q^\mathrm{visc}$ is given in \citep{Narayan_Yi_original} as
\begin{equation}
    q^\mathrm{visc} = \frac{3 \epsilon' \rho v_R c_s^2}{2R} = 0.08\epsilon' \, m_{\mathrm{BH},9}^{-2} \, \dot{m}_{-3} \, r^{-4}  \, \, \mathrm{erg} \, \mathrm{cm}^{-3} \, \mathrm{s}^{-1}
\end{equation}
where $\epsilon' = (5/3 - \gamma')/(\gamma' - 1)$, with $\gamma' = (32 - 24\beta - 3\beta^2)/(24 - 21\beta) $.

For the Coulomb interaction heating rate per unit volume $q^\mathrm{ie}$, from \cite{rates_hotplasma}, assuming $n_e \approx n_p$, we use the approximation from \cite{Mahadevan1997}.

\begin{equation}
    q^\mathrm{ie} = 5.61\times 10^{-32} \frac{n_e^2(T_i - T_e)}{K_2(1/\theta_e)}\left(\frac{\theta_e\theta_i}{\theta_i(\theta_e + \theta_i}\right)^{1/2}\left[\frac{2(\theta_e + \theta_i)^2 + 1 + 2(\theta_e + \theta_i}{(\theta_e + \theta_i)}\right]e^{-1/\theta_e}  \, \, \mathrm{erg} \, \mathrm{cm}^{-3} \, \mathrm{s}^{-1} 
\end{equation}

The total heating rate is then given by
\begin{equation}
    q^{e+} =q^\mathrm{ie} + \delta_e q^\mathrm{visc} \label{eq:Qeplus}
\end{equation}

\section{Truncated thin disc} \label{appendix disc}

For completeness, we compute the spectrum from an outer disc, such that the inner radius of the disc is equal to the outer radius of the accretion flow $r_\mathrm{tr} = r_\mathrm{max}$.
The emission is characterized by the sum of blackbody spectra with temperature 
\begin{equation}
    T_\mathrm{disc}(R) = \left( \frac{G\, M_\mathrm{BH}\, \dot{M}}{8\pi \, R^3 \, \sigma_\mathrm{B}}\left[ 1 - \left( \frac{R_\mathrm{tr}}{R} \right)^{1/2}\right] \right)^{1/4}
\end{equation}
The emission is then given by 
\begin{equation}
    F_{\nu, \mathrm{disc}} = \frac{4\pi h \cos{\theta} \nu^3}{c^2 D^2}\int_{R_\mathrm{tr}}^{R_\mathrm{max,disc}} \, \frac{R \mathrm{d}R}{e^{h\nu/k_\mathrm{B}T_\mathrm{disc}(R)} - 1}
\end{equation}
where we have set an outer radius of $\sim 3\times 10^6$ but this parameter has a poor influence on the spectrum, given the large truncation radius.

\bibliography{references}{}

\begin{thebibliography}{}
\expandafter\ifx\csname natexlab\endcsname\relax\def\natexlab#1{#1}\fi
\providecommand{\url}[1]{\href{#1}{#1}}
\providecommand{\dodoi}[1]{doi:~\href{http://doi.org/#1}{\nolinkurl{#1}}}
\providecommand{\doeprint}[1]{\href{http://ascl.net/#1}{\nolinkurl{http://ascl.net/#1}}}
\providecommand{\doarXiv}[1]{\href{https://arxiv.org/abs/#1}{\nolinkurl{https://arxiv.org/abs/#1}}}

\bibitem[{{Aab} {et~al.}(2019){Aab}, {Abreu}, {Aglietta}, {Albuquerque},
  {Albury}, {Allekotte}, {Almela}, {Alvarez Castillo}, {Alvarez-Mu{\~n}iz},
  {Anastasi}, {Anchordoqui}, {Andrada}, {Andringa}, {Aramo}, {Asorey}, {Assis},
  {Avila}, {Badescu}, {Bakalova}, {Balaceanu}, {Barbato}, {Barreira Luz},
  {Baur}, {Becker}, {Bellido}, {Berat}, {Bertaina}, {Bertou}, {Biermann},
  {Biteau}, {Blanco}, {Blazek}, {Bleve}, {Boh{\'a}{\v{c}}ov{\'a}}, {Boncioli},
  {Bonifazi}, {Borodai}, {Botti}, {Brack}, {Bretz}, {Bridgeman}, {Briechle},
  {Buchholz}, {Bueno}, {Buitink}, {Buscemi}, {Caballero-Mora}, {Caccianiga},
  {Calcagni}, {Cancio}, {Canfora}, {Caracas}, {Carceller}, {Caruso},
  {Castellina}, {Catalani}, {Cataldi}, {Cazon}, {Cerda}, {Chinellato}, {Choi},
  {Chudoba}, {Chytka}, {Clay}, {Cobos Cerutti}, {Colalillo}, {Coleman},
  {Coluccia}, {Concei{\c{c}}{\~a}o}, {Condorelli}, {Consolati}, {Contreras},
  {Convenga}, {Cooper}, {Coutu}, {Covault}, {Daniel}, {Dasso}, {Daumiller},
  {Dawson}, {Day}, {de Almeida}, {de Jong}, {De Mauro}, {de Mello Neto}, {De
  Mitri}, {de Oliveira}, {de Souza}, {Debatin}, {del R{\'\i}o}, {Deligny},
  {Dhital}, {Di Matteo}, {D{\'\i}az Castro}, {Dobrigkeit}, {D'Olivo},
  {Dorosti}, {dos Anjos}, {Dova}, {Dundovic}, {Ebr}, {Engel}, {Erdmann},
  {Escobar}, {Etchegoyen}, {Falcke}, {Farmer}, {Farrar}, {Fauth}, {Fazzini},
  {Feldbusch}, {Fenu}, {Ferreyro}, {Figueira}, {Filip{\v{c}}i{\v{c}}},
  {Freire}, {Fujii}, {Fuster}, {Garc{\'\i}a}, {Gemmeke}, {Gesualdi},
  {Gherghel-Lascu}, {Ghia}, {Giaccari}, {Giammarchi}, {Giller}, {G{\l}as},
  {Glombitza}, {Gobbi}, {Golup}, {G{\'o}mez Berisso}, {G{\'o}mez Vitale},
  {Gongora}, {Gonz{\'a}lez}, {Goos}, {G{\'o}ra}, {Gorgi}, {Gottowik}, {Grubb},
  {Guarino}, {Guedes}, {Guido}, {Hahn}, {Halliday}, {Hampel}, {Hansen},
  {Harari}, {Harrison}, {Harvey}, {Haungs}, {Hebbeker}, {Heck}, {Heimann},
  {Hill}, {Hojvat}, {Holt}, {Homola}, {H{\"o}randel}, {Horvath},
  {Hrabovsk{\'y}}, {Huege}, {Hulsman}, {Insolia}, {Isar}, {Johnsen}, {Jurysek},
  {K{\"a}{\"a}p{\"a}}, {Kampert}, {Keilhauer}, {Kemmerich}, {Kemp}, {Klages},
  {Kleifges}, {Kleinfeller}, {Kuempel}, {Kukec Mezek}, {Kuotb Awad}, {Lago},
  {LaHurd}, {Lang}, {Legumina}, {Leigui de Oliveira}, {Lenok},
  {Letessier-Selvon}, {Lhenry-Yvon}, {Lippmann}, {Lo Presti}, {Lopes},
  {L{\'o}pez}, {L{\'o}pez Casado}, {Lorek}, {Luce}, {Lucero}, {Malacari},
  {Mancarella}, {Mandat}, {Manning}, {Manshanden}, {Mantsch}, {Mariazzi},
  {Mari{\c{s}}}, {Marsella}, {Martello}, {Martinez}, {Mart{\'\i}nez Bravo},
  {Mastrodicasa}, {Mathes}, {Mathys}, {Matthews}, {Matthiae}, {Mayotte},
  {Mazur}, {Medina-Tanco}, {Melo}, {Menshikov}, {Merenda}, {Michal},
  {Micheletti}, {Miramonti}, {Mockler}, {Mollerach}, {Montanet}, {Morello},
  {Morlino}, {Mostaf{\'a}}, {M{\"u}ller}, {Muller}, {M{\"u}ller}, {Mussa},
  {Namasaka}, {Nellen}, {Niculescu-Oglinzanu}, {Niechciol}, {Nitz}, {Nosek},
  {Novotny}, {No{\v{z}}ka}, {Nucita}, {N{\'u}{\~n}ez}, {Olinto}, {Palatka},
  {Pallotta}, {Panetta}, {Papenbreer}, {Parente}, {Parra}, {Pech}, {Pedreira},
  {P{\c{e}}kala}, {Pelayo}, {Pe{\~n}a-Rodriguez}, {Pereira}, {Perlin},
  {Perrone}, {Peters}, {Petrera}, {Phuntsok}, {Pierog}, {Pimenta},
  {Pirronello}, {Platino}, {Poh}, {Pont}, {Porowski}, {Pothast}, {Prado},
  {Privitera}, {Prouza}, {Puyleart}, {Querchfeld}, {Quinn}, {Ramos-Pollan},
  {Rautenberg}, {Ravignani}, {Reininghaus}, {Ridky}, {Riehn}, {Risse},
  {Ristori}, {Rizi}, {Rodrigues de Carvalho}, {Rodriguez Rojo}, {Roncoroni},
  {Roth}, {Roulet}, {Rovero}, {Ruehl}, {Saffi}, {Saftoiu}, {Salamida},
  {Salazar}, {Salina}, {Sanabria Gomez}, {S{\'a}nchez}, {Santos}, {Santos},
  {Sarazin}, {Sarmento}, {Sarmiento-Cano}, {Sato}, {Savina}, {Schauer},
  {Scherini}, {Schieler}, {Schimassek}, {Schimp}, {Schl{\"u}ter}, {Schmidt},
  {Scholten}, {Schov{\'a}nek}, {Schr{\"o}der}, {Schr{\"o}der}, {Schumacher},
  {Sciutto}, {Scornavacche}, {Shellard}, {Sigl}, {Silli}, {Sima},
  {{\v{S}}m{\'\i}da}, {Snow}, {Sommers}, {Soriano}, {Souchard}, {Squartini},
  {Stadelmaier}, {Stanca}, {Stani{\v{c}}}, {Stasielak}, {Stassi},
  {Stolpovskiy}, {Streich}, {Su{\'a}rez-Dur{\'a}n}, {Sudholz},
  {Suomij{\"a}rvi}, {Supanitsky}, {{\v{S}}up{\'\i}k}, {Szadkowski}, {Taboada},
  {Taborda}, {Tapia}, {Timmermans}, {Tobiska}, {Todero Peixoto}, {Tom{\'e}},
  {Torralba Elipe}, {Travaini}, {Travnicek}, {Trini}, {Tueros}, {Ulrich},
  {Unger}, {Urban}, {Vald{\'e}s Galicia}, {Vali{\~n}o}, {Valore}, {van
  Bodegom}, {van den Berg}, {van Vliet}, {Varela}, {Vargas C{\'a}rdenas},
  {V{\'a}squez-Ram{\'\i}rez}, {Veberi{\v{c}}}, {Ventura}, {Vergara Quispe},
  {Verzi}, {Vicha}, {Villase{\~n}or}, {Vink}, {Vorobiov}, {Wahlberg}, {Watson},
  {Weber}, {Weindl}, {Wiede{\'n}ski}, {Wiencke}, {Wilczy{\'n}ski}, {Winchen},
  {Wirtz}, {Wittkowski}, {Wundheiler}, {Yang}, {Yushkov}, {Zas}, {Zavrtanik},
  {Zavrtanik}, {Zehrer}, {Zepeda}, {Zimmermann}, {Ziolkowski}, \&
  {Zuccarello}}]{PierreAuger}
{Aab}, A., {Abreu}, P., {Aglietta}, M., {et~al.} 2019, \jcap, 2019, 004,
  \dodoi{10.1088/1475-7516/2019/11/004}

\bibitem[{{Aartsen} {et~al.}(2017){Aartsen}, {Abraham}, {Ackermann}, {Adams},
  {Aguilar}, {Ahlers}, {Ahrens}, {Altmann}, {Andeen}, {Anderson}, {Ansseau},
  {Anton}, {Archinger}, {Arg{\"u}elles}, {Auffenberg}, {Axani}, {Bai},
  {Barwick}, {Baum}, {Bay}, {Beatty}, {Becker Tjus}, {Becker}, {BenZvi},
  {Berley}, {Bernardini}, {Bernhard}, {Besson}, {Binder}, {Bindig}, {Bissok},
  {Blaufuss}, {Blot}, {Bohm}, {B{\"o}rner}, {Bos}, {Bose}, {B{\"o}ser},
  {Botner}, {Braun}, {Brayeur}, {Bretz}, {Bron}, {Burgman}, {Carver}, {Casier},
  {Cheung}, {Chirkin}, {Christov}, {Clark}, {Classen}, {Coenders}, {Collin},
  {Conrad}, {Cowen}, {Cross}, {Day}, {de Andr{\'e}}, {De Clercq}, {del Pino
  Rosendo}, {Dembinski}, {De Ridder}, {Desiati}, {de Vries}, {de Wasseige}, {de
  With}, {DeYoung}, {D{\'\i}az-V{\'e}lez}, {di Lorenzo}, {Dujmovic}, {Dumm},
  {Dunkman}, {Eberhardt}, {Ehrhardt}, {Eichmann}, {Eller}, {Euler}, {Evenson},
  {Fahey}, {Fazely}, {Feintzeig}, {Felde}, {Filimonov}, {Finley}, {Flis},
  {F{\"o}sig}, {Franckowiak}, {Friedman}, {Fuchs}, {Gaisser}, {Gallagher},
  {Gerhardt}, {Ghorbani}, {Giang}, {Gladstone}, {Glauch}, {Gl{\"u}senkamp},
  {Goldschmidt}, {Golup}, {Gonzalez}, {Grant}, {Griffith}, {Haack}, {Haj
  Ismail}, {Hallgren}, {Halzen}, {Hansen}, {Hansmann}, {Hanson}, {Hebecker},
  {Heereman}, {Helbing}, {Hellauer}, {Hickford}, {Hignight}, {Hill}, {Hoffman},
  {Hoffmann}, {Holzapfel}, {Hoshina}, {Huang}, {Huber}, {Hultqvist}, {In},
  {Ishihara}, {Jacobi}, {Japaridze}, {Jeong}, {Jero}, {Jones}, {Jurkovic},
  {Kappes}, {Karg}, {Karle}, {Katz}, {Kauer}, {Keivani}, {Kelley},
  {Kheirandish}, {Kim}, {Kintscher}, {Kiryluk}, {Kittler}, {Klein}, {Kohnen},
  {Koirala}, {Kolanoski}, {Konietz}, {K{\"o}pke}, {Kopper}, {Kopper},
  {Koskinen}, {Kowalski}, {Krings}, {Kroll}, {Kr{\"u}ckl}, {Kr{\"u}ger},
  {Kunnen}, {Kunwar}, {Kurahashi}, {Kuwabara}, {Labare}, {Lanfranchi},
  {Larson}, {Lauber}, {Lennarz}, {Lesiak-Bzdak}, {Leuermann}, {Lu},
  {L{\"u}nemann}, {Madsen}, {Maggi}, {Mahn}, {Mancina}, {Mandelartz},
  {Maruyama}, {Mase}, {Maunu}, {McNally}, {Meagher}, {Medici}, {Meier}, {Meli},
  {Menne}, {Merino}, {Meures}, {Miarecki}, {Mohrmann}, {Montaruli}, {Moulai},
  {Nahnhauer}, {Naumann}, {Neer}, {Niederhausen}, {Nowicki}, {Nygren},
  {Obertacke Pollmann}, {Olivas}, {O'Murchadha}, {Palczewski}, {Pandya},
  {Pankova}, {Peiffer}, {Penek}, {Pepper}, {P{\'e}rez de los Heros}, {Pieloth},
  {Pinat}, {Price}, {Przybylski}, {Quinnan}, {Raab}, {R{\"a}del}, {Rameez},
  {Rawlins}, {Reimann}, {Relethford}, {Relich}, {Resconi}, {Rhode}, {Richman},
  {Riedel}, {Robertson}, {Rongen}, {Rott}, {Ruhe}, {Ryckbosch}, {Rysewyk},
  {Sabbatini}, {Sanchez Herrera}, {Sandrock}, {Sandroos}, {Sarkar},
  {Satalecka}, {Schlunder}, {Schmidt}, {Schoenen}, {Sch{\"o}neberg},
  {Schumacher}, {Seckel}, {Seunarine}, {Soldin}, {Song}, {Spiczak}, {Spiering},
  {Stanev}, {Stasik}, {Stettner}, {Steuer}, {Stezelberger}, {Stokstad},
  {St{\"o}ssl}, {Str{\"o}m}, {Strotjohann}, {Sullivan}, {Sutherland},
  {Taavola}, {Taboada}, {Tatar}, {Tenholt}, {Ter-Antonyan}, {Terliuk},
  {Te{\v{s}}i{\'c}}, {Tilav}, {Toale}, {Tobin}, {Toscano}, {Tosi},
  {Tselengidou}, {Turcati}, {Unger}, {Usner}, {Vandenbroucke}, {van
  Eijndhoven}, {Vanheule}, {van Rossem}, {van Santen}, {Veenkamp}, {Vehring},
  {Voge}, {Vogel}, {Vraeghe}, {Walck}, {Wallace}, {Wallraff}, {Wandkowsky},
  {Weaver}, {Weiss}, {Wendt}, {Westerhoff}, {Whelan}, {Wickmann}, {Wiebe},
  {Wiebusch}, {Wille}, {Williams}, {Wills}, {Wolf}, {Wood}, {Woolsey},
  {Woschnagg}, {Xu}, {Xu}, {Xu}, {Yanez}, {Yodh}, {Yoshida}, {Zoll}, \&
  {IceCube Collaboration}}]{IceCube}
{Aartsen}, M.~G., {Abraham}, K., {Ackermann}, M., {et~al.} 2017, \apj, 835,
  151, \dodoi{10.3847/1538-4357/835/2/151}

\bibitem[{{Abramowicz} {et~al.}(1995){Abramowicz}, {Chen}, {Kato}, {Lasota}, \&
  {Regev}}]{adafintro}
{Abramowicz}, M.~A., {Chen}, X., {Kato}, S., {Lasota}, J.-P., \& {Regev}, O.
  1995, \apjl, 438, L37, \dodoi{10.1086/187709}

\bibitem[{{Asada} \& {Nakamura}(2012)}]{bondiradius}
{Asada}, K., \& {Nakamura}, M. 2012, \apjl, 745, L28,
  \dodoi{10.1088/2041-8205/745/2/L28}

\bibitem[{{Biermann} {et~al.}(2000){Biermann}, {Ahn}, {Medina-Tanco}, \&
  {Stanev}}]{UHECR_M87}
{Biermann}, P.~L., {Ahn}, E.-J., {Medina-Tanco}, G., \& {Stanev}, T. 2000,
  Nuclear Physics B Proceedings Supplements, 87, 417,
  \dodoi{10.1016/S0920-5632(00)00708-8}

\bibitem[{{Bisnovatyi-Kogan} \& {Ruzmaikin}(1976)}]{firstMAD}
{Bisnovatyi-Kogan}, G.~S., \& {Ruzmaikin}, A.~A. 1976, \apss, 42, 401,
  \dodoi{10.1007/BF01225967}

\bibitem[{{Blandford} \& {Begelman}(1999)}]{Blandford_Begelman_massloss}
{Blandford}, R.~D., \& {Begelman}, M.~C. 1999, \mnras, 303, L1,
  \dodoi{10.1046/j.1365-8711.1999.02358.x}

\bibitem[{{Blandford} \& {K{\"o}nigl}(1979{\natexlab{a}})}]{BlandfordSteady}
{Blandford}, R.~D., \& {K{\"o}nigl}, A. 1979{\natexlab{a}}, \apj, 232, 34,
  \dodoi{10.1086/157262}

\bibitem[{{Blandford} \&
  {K{\"o}nigl}(1979{\natexlab{b}})}]{Blandford_prediction_SSA}
---. 1979{\natexlab{b}}, \apj, 232, 34, \dodoi{10.1086/157262}

\bibitem[{{Blandford} \& {Znajek}(1977)}]{Blandford_Znajek}
{Blandford}, R.~D., \& {Znajek}, R.~L. 1977, \mnras, 179, 433,
  \dodoi{10.1093/mnras/179.3.433}

\bibitem[{{Chael} {et~al.}(2019){Chael}, {Narayan}, \&
  {Johnson}}]{alpha_visco_mad}
{Chael}, A., {Narayan}, R., \& {Johnson}, M.~D. 2019, \mnras, 486, 2873,
  \dodoi{10.1093/mnras/stz988}

\bibitem[{{Chael} {et~al.}(2018){Chael}, {Rowan}, {Narayan}, {Johnson}, \&
  {Sironi}}]{beta_values1}
{Chael}, A., {Rowan}, M., {Narayan}, R., {Johnson}, M., \& {Sironi}, L. 2018,
  \mnras, 478, 5209, \dodoi{10.1093/mnras/sty1261}

\bibitem[{{Coppi} \& {Blandford}(1990)}]{Coppi_Blandford_scattering}
{Coppi}, P.~S., \& {Blandford}, R.~D. 1990, \mnras, 245, 453

\bibitem[{{de Gasperin} {et~al.}(2012){de Gasperin}, {Orr{\'u}}, {Murgia},
  {Merloni}, {Falcke}, {Beck}, {Beswick}, {B{\^\i}rzan}, {Bonafede},
  {Br{\"u}ggen}, {Brunetti}, {Chy{\.z}y}, {Conway}, {Croston}, {En{\ss}lin},
  {Ferrari}, {Heald}, {Heidenreich}, {Jackson}, {Macario}, {McKean}, {Miley},
  {Morganti}, {Offringa}, {Pizzo}, {Rafferty}, {R{\"o}ttgering}, {Shulevski},
  {Steinmetz}, {Tasse}, {van der Tol}, {van Driel}, {van Weeren}, {van
  Zwieten}, {Alexov}, {Anderson}, {Asgekar}, {Avruch}, {Bell}, {Bell},
  {Bentum}, {Bernardi}, {Best}, {Breitling}, {Broderick}, {Butcher}, {Ciardi},
  {Dettmar}, {Eisloeffel}, {Frieswijk}, {Gankema}, {Garrett}, {Gerbers},
  {Griessmeier}, {Gunst}, {Hassall}, {Hessels}, {Hoeft}, {Horneffer},
  {Karastergiou}, {K{\"o}hler}, {Koopman}, {Kuniyoshi}, {Kuper}, {Maat},
  {Mann}, {Mevius}, {Mulcahy}, {Munk}, {Nijboer}, {Noordam}, {Paas}, {Pandey},
  {Pandey}, {Polatidis}, {Reich}, {Schoenmakers}, {Sluman}, {Smirnov}, {Sobey},
  {Stappers}, {Swinbank}, {Tagger}, {Tang}, {van Bemmel}, {van Cappellen}, {van
  Duin}, {van Haarlem}, {van Leeuwen}, {Vermeulen}, {Vocks}, {White}, {Wise},
  {Wucknitz}, \& {Zarka}}]{Jet_power}
{de Gasperin}, F., {Orr{\'u}}, E., {Murgia}, M., {et~al.} 2012, \aap, 547, A56,
  \dodoi{10.1051/0004-6361/201220209}

\bibitem[{{Di Matteo} {et~al.}(2003){Di Matteo}, {Allen}, {Fabian}, {Wilson},
  \& {Young}}]{DiMatteoM87}
{Di Matteo}, T., {Allen}, S.~W., {Fabian}, A.~C., {Wilson}, A.~S., \& {Young},
  A.~J. 2003, \apj, 582, 133, \dodoi{10.1086/344504}

\bibitem[{{Dom{\'\i}nguez} {et~al.}(2011){Dom{\'\i}nguez}, {Primack},
  {Rosario}, {Prada}, {Gilmore}, {Faber}, {Koo}, {Somerville},
  {P{\'e}rez-Torres}, {P{\'e}rez-Gonz{\'a}lez}, {Huang}, {Davis},
  {Guhathakurta}, {Barmby}, {Conselice}, {Lozano}, {Newman}, \&
  {Cooper}}]{EBL_dominguez}
{Dom{\'\i}nguez}, A., {Primack}, J.~R., {Rosario}, D.~J., {et~al.} 2011,
  \mnras, 410, 2556, \dodoi{10.1111/j.1365-2966.2010.17631.x}

\bibitem[{{EHT MWL Science Working Group} {et~al.}(2021){EHT MWL Science
  Working Group}, {Algaba}, {Anczarski}, {Asada}, {Balokovi{\'c}}, {Chandra},
  {Cui}, {Falcone}, {Giroletti}, {Goddi}, {Hada}, {Haggard}, {Jorstad}, {Kaur},
  {Kawashima}, {Keating}, {Kim}, {Kino}, {Komossa}, {Kravchenko}, {Krichbaum},
  {Lee}, {Lu}, {Lucchini}, {Markoff}, {Neilsen}, {Nowak}, {Park}, {Principe},
  {Ramakrishnan}, {Reynolds}, {Sasada}, {Savchenko}, {Williamson}, {Event
  Horizon Telescope Collaboration}, {Akiyama}, {Alberdi}, {Alef}, {Anantua},
  {Azulay}, {Baczko}, {Ball}, {Barrett}, {Bintley}, {Benson}, {Blackburn},
  {Blundell}, {Boland}, {Bouman}, {Bower}, {Boyce}, {Bremer}, {Brinkerink},
  {Brissenden}, {Britzen}, {Broderick}, {Broguiere}, {Bronzwaer}, {Byun},
  {Carlstrom}, {Chael}, {Chan}, {Chatterjee}, {Chatterjee}, {Chen}, {Chen},
  {Chesler}, {Cho}, {Christian}, {Conway}, {Cordes}, {Crawford}, {Crew},
  {Cruz-Osorio}, {Davelaar}, {de Laurentis}, {Deane}, {Dempsey}, {Desvignes},
  {Dexter}, {Doeleman}, {Eatough}, {Falcke}, {Farah}, {Fish}, {Fomalont},
  {Ford}, {Fraga-Encinas}, {Friberg}, {Fromm}, {Fuentes}, {Galison}, {Gammie},
  {Garc{\'\i}a}, {Gentaz}, {Georgiev}, {Gold}, {G{\'o}mez}, {G{\'o}mez-Ruiz},
  {Gu}, {Gurwell}, {Hecht}, {Hesper}, {Ho}, {Ho}, {Honma}, {Huang}, {Huang},
  {Hughes}, {Ikeda}, {Inoue}, {Issaoun}, {James}, {Jannuzi}, {Janssen},
  {Jeter}, {Jiang}, {Jim{\'e}nez-Rosales}, {Johnson}, {Jung}, {Karami},
  {Karuppusamy}, {Kettenis}, {Kim}, {Kim}, {Kim}, {Koay}, {Kofuji}, {Koch},
  {Koyama}, {Kramer}, {Kramer}, {Kuo}, {Lauer}, {Levis}, {Li}, {Li},
  {Lindqvist}, {Lico}, {Lindahl}, {Liu}, {Liu}, {Liuzzo}, {Lo}, {Lobanov},
  {Loinard}, {Lonsdale}, {MacDonald}, {Mao}, {Marchili}, {Marrone}, {Marscher},
  {Mart{\'\i}-Vidal}, {Matsushita}, {Matthews}, {Medeiros}, {Menten}, {Mizuno},
  {Mizuno}, {Moran}, {Moriyama}, {Moscibrodzka}, {M{\"u}ller}, {Musoke},
  {Mej{\'\i}as}, {Nagai}, {Nagar}, {Nakamura}, {Narayan}, {Narayanan},
  {Natarajan}, {Nathanail}, {Neri}, {Ni}, {Noutsos}, {Okino}, {Olivares},
  {Ortiz-Le{\'o}n}, {Oyama}, {{\"O}zel}, {Palumbo}, {Patel}, {Pen}, {Pesce},
  {Pi{\'e}tu}, {Plambeck}, {Popstefanija}, {Porth}, {P{\"o}tzl}, {Prather},
  {Preciado-L{\'o}pez}, {Psaltis}, {Pu}, {Rao}, {Rawlings}, {Raymond},
  {Rezzolla}, {Ricarte}, {Ripperda}, {Roelofs}, {Rogers}, {Ros}, {Rose},
  {Roshanineshat}, {Rottmann}, {Roy}, {Ruszczyk}, {Rygl}, {S{\'a}nchez},
  {S{\'a}nchez-Arguelles}, {Savolainen}, {Schloerb}, {Schuster}, {Shao},
  {Shen}, {Small}, {Sohn}, {Soohoo}, {Sun}, {Tazaki}, {Tetarenko}, {Tiede},
  {Tilanus}, {Titus}, {Toma}, {Torne}, {Trent}, {Traianou}, {Trippe}, {van
  Bemmel}, {van Langevelde}, {van Rossum}, {Wagner}, {Ward-Thompson}, {Wardle},
  {Weintroub}, {Wex}, {Wharton}, {Wielgus}, {Wong}, {Wu}, {Yoon}, {Young},
  {Young}, {Younsi}, {Yuan}, {Yuan}, {Zensus}, {Zhao}, {Zhao}, {Fermi Large
  Area Telescope Collaboration}, {Principe}, {Giroletti}, {D'Ammando},
  {Orienti}, {H.~E.~S.~S. Collaboration}, {Abdalla}, {Adam}, {Aharonian},
  {Benkhali}, {Ang{\"u}ner}, {Arcaro}, {Armand}, {Armstrong}, {Ashkar},
  {Backes}, {Baghmanyan}, {Barbosa Martins}, {Barnacka}, {Barnard},
  {Becherini}, {Berge}, {Bernl{\"o}hr}, {Bi}, {B{\"o}ttcher}, {Boisson},
  {Bolmont}, {de Lavergne}, {Breuhaus}, {Brun}, {Brun}, {Bryan}, {B{\"u}chele},
  {Bulik}, {Bylund}, {Caroff}, {Carosi}, {Casanova}, {Chand}, {Chen}, {Cotter},
  {Cury{\l}o}, {Damascene Mbarubucyeye}, {Davids}, {Davies}, {Deil}, {Devin},
  {Dewilt}, {Dirson}, {Djannati-Ata{\"\i}}, {Dmytriiev}, {Donath},
  {Doroshenko}, {Duffy}, {Dyks}, {Egberts}, {Eichhorn}, {Einecke}, {Emery},
  {Ernenwein}, {Feijen}, {Fegan}, {Fiasson}, {de Clairfontaine}, {Fontaine},
  {Funk}, {F{\"u}{\ss}ling}, {Gabici}, {Gallant}, {Giavitto}, {Giunti},
  {Glawion}, {Glicenstein}, {Gottschall}, {Grondin}, {Hahn}, {Haupt},
  {Hermann}, {Hinton}, {Hofmann}, {Hoischen}, {Holch}, {Holler}, {H{\"o}rbe},
  {Horns}, {Huber}, {Jamrozy}, {Jankowsky}, {Jankowsky}, {Jardin-Blicq},
  {Joshi}, {Jung-Richardt}, {Kasai}, {Kastendieck}, {Katarzy{\'n}ski}, {Katz},
  {Khangulyan}, {Kh{\'e}lifi}, {Klepser}, {Klu{\'z}niak}, {Komin}, {Konno},
  {Kosack}, {Kostunin}, {Kreter}, {Lamanna}, {Lemi{\`e}re}, {Lemoine-Goumard},
  {Lenain}, {Levy}, {Lohse}, {Lypova}, {Mackey}, {Majumdar}, {Malyshev},
  {Malyshev}, {Marandon}, {Marchegiani}, {Marcowith}, {Mares},
  {Mart{\'\i}-Devesa}, {Marx}, {Maurin}, {Meintjes}, {Meyer}, {Moderski},
  {Mohamed}, {Mohrmann}, {Montanari}, {Moore}, {Morris}, {Moulin}, {Muller},
  {Murach}, {Nakashima}, {Nayerhoda}, {de Naurois}, {Ndiyavala},
  {Niederwanger}, {Niemiec}, {Oakes}, {O'Brien}, {Odaka}, {Ohm},
  {Olivera-Nieto}, {de Ona Wilhelmi}, {Ostrowski}, {Panter}, {Panny},
  {Parsons}, {Peron}, {Peyaud}, {Piel}, {Pita}, {Poireau}, {Noel}, {Prokhorov},
  {Prokoph}, {P{\"u}hlhofer}, {Punch}, {Quirrenbach}, {Rauth}, {Reichherzer},
  {Reimer}, {Reimer}, {Remy}, {Renaud}, {Rieger}, {Rinchiuso}, {Romoli},
  {Rowell}, {Rudak}, {Ruiz-Velasco}, {Sahakian}, {Sailer}, {Sanchez},
  {Santangelo}, {Sasaki}, {Scalici}, {Schutte}, {Schwanke}, {Schwemmer},
  {Seglar-Arroyo}, {Senniappan}, {Seyffert}, {Shafi}, {Shiningayamwe},
  {Simoni}, {Sinha}, {Sol}, {Specovius}, {Spencer}, {Spir-Jacob}, {Stawarz},
  {Sun}, {Steenkamp}, {Stegmann}, {Steinmassl}, {Steppa}, {Takahashi},
  {Tavernier}, {Taylor}, {Terrier}, {Tiziani}, {Tluczykont}, {Tomankova},
  {Trichard}, {Tsirou}, {Tuffs}, {Uchiyama}, {van der Walt}, {van Eldik}, {van
  Rensburg}, {van Soelen}, {Vasileiadis}, {Veh}, {Venter}, {Vincent}, {Vink},
  {V{\"o}lk}, {Vuillaume}, {Wadiasingh}, {Wagner}, {Watson}, {Werner}, {White},
  {Wierzcholska}, {Wong}, {Yusafzai}, {Zacharias}, {Zanin}, {Zargaryan},
  {Zdziarski}, {Zech}, {Zhu}, {Zorn}, {Zouari}, {{\.Z}ywucka}, {MAGIC
  Collaboration}, {Acciari}, {Ansoldi}, {Antonelli}, {Engels}, {Artero},
  {Asano}, {Baack}, {Babi{\'c}}, {Baquero}, {de Almeida}, {Barrio}, {Becerra
  Gonz{\'a}lez}, {Bednarek}, {Bellizzi}, {Bernardini}, {Bernardos}, {Berti},
  {Besenrieder}, {Bhattacharyya}, {Bigongiari}, {Biland}, {Blanch}, {Bonnoli},
  {Bo{\v{s}}njak}, {Busetto}, {Carosi}, {Ceribella}, {Cerruti}, {Chai},
  {Chilingarian}, {Cikota}, {Colak}, {Colombo}, {Contreras}, {Cortina},
  {Covino}, {D'Amico}, {D'Elia}, {da Vela}, {Dazzi}, {de Angelis}, {de Lotto},
  {Delfino}, {Delgado}, {Delgado Mendez}, {Depaoli}, {di Pierro}, {di Venere},
  {Do Souto Espi{\~n}eira}, {Dominis Prester}, {Donini}, {Dorner}, {Doro},
  {Elsaesser}, {Ramazani}, {Fattorini}, {Ferrara}, {Fonseca}, {Font}, {Fruck},
  {Fukami}, {Garc{\'\i}a L{\'o}pez}, {Garczarczyk}, {Gasparyan}, {Gaug},
  {Giglietto}, {Giordano}, {Gliwny}, {Godinovi{\'c}}, {Green}, {Green},
  {Hadasch}, {Hahn}, {Heckmann}, {Herrera}, {Hoang}, {Hrupec}, {H{\"u}tten},
  {Inada}, {Inoue}, {Ishio}, {Iwamura}, {Jim{\'e}nez}, {Jormanainen}, {Jouvin},
  {Kajiwara}, {Karjalainen}, {Kerszberg}, {Kobayashi}, {Kubo}, {Kushida},
  {Lamastra}, {Lelas}, {Leone}, {Lindfors}, {Lombardi}, {Longo},
  {L{\'o}pez-Coto}, {L{\'o}pez-Moya}, {L{\'o}pez-Oramas}, {Loporchio}, {Machado
  de Oliveira Fraga}, {Maggio}, {Majumdar}, {Makariev}, {Mallamaci}, {Maneva},
  {Manganaro}, {Mannheim}, {Maraschi}, {Mariotti}, {Mart{\'\i}nez}, {Mazin},
  {Menchiari}, {Mender}, {Mi{\'c}anovi{\'c}}, {Miceli}, {Miener}, {Minev},
  {Miranda}, {Mirzoyan}, {Molina}, {Moralejo}, {Morcuende}, {Moreno},
  {Moretti}, {Neustroev}, {Nigro}, {Nilsson}, {Nishijima}, {Noda}, {Nozaki},
  {Ohtani}, {Oka}, {Otero-Santos}, {Paiano}, {Palatiello}, {Paneque},
  {Paoletti}, {Paredes}, {Pavleti{\'c}}, {Pe{\~n}il}, {Perennes}, {Persic},
  {Moroni}, {Prandini}, {Priyadarshi}, {Puljak}, {Rhode}, {Rib{\'o}}, {Rico},
  {Righi}, {Rugliancich}, {Saha}, {Sahakyan}, {Saito}, {Sakurai}, {Satalecka},
  {Saturni}, {Schleicher}, {Schmidt}, {Schweizer}, {Sitarek},
  {{\v{S}}nidari{\'c}}, {Sobczynska}, {Spolon}, {Stamerra}, {Strom}, {Strzys},
  {Suda}, {Suri{\'c}}, {Takahashi}, {Tavecchio}, {Temnikov}, {Terzi{\'c}},
  {Teshima}, {Tosti}, {Truzzi}, {Tutone}, {Ubach}, {van Scherpenberg}, {Vanzo},
  {Vazquez Acosta}, {Ventura}, {Verguilov}, {Vigorito}, {Vitale}, {Vovk},
  {Will}, {Wunderlich}, {Zari{\'c}}, {VERITAS Collaboration}, {Adams},
  {Benbow}, {Brill}, {Capasso}, {Christiansen}, {Chromey}, {Daniel}, {Errando},
  {Farrell}, {Feng}, {Finley}, {Fortson}, {Furniss}, {Gent}, {Giuri}, {Hassan},
  {Hervet}, {Holder}, {Hughes}, {Humensky}, {Jin}, {Kaaret}, {Kertzman},
  {Kieda}, {Kumar}, {Lang}, {Lundy}, {Maier}, {Moriarty}, {Mukherjee}, {Nieto},
  {Nievas-Rosillo}, {O'Brien}, {Ong}, {Otte}, {Patel}, {Pfrang}, {Pohl},
  {Prado}, {Pueschel}, {Quinn}, {Ragan}, {Reynolds}, {Ribeiro}, {Richards},
  {Roache}, {Rulten}, {Ryan}, {Santander}, {Sembroski}, {Shang}, {Weinstein},
  {Williams}, {Williamson}, {Eavn Collaboration}, {Hirota}, {Cui}, {Niinuma},
  {Ro}, {Sakai}, {Sawada-Satoh}, {Wajima}, {Wang}, {Liu}, \&
  {Yonekura}}]{EHTpaper}
{EHT MWL Science Working Group}, {Algaba}, J.~C., {Anczarski}, J., {et~al.}
  2021, \apjl, 911, L11, \dodoi{10.3847/2041-8213/abef71}

\bibitem[{{Esin} {et~al.}(1996){Esin}, {Narayan}, {Ostriker}, \&
  {Yi}}]{Esin_compton_enhancement}
{Esin}, A.~A., {Narayan}, R., {Ostriker}, E., \& {Yi}, I. 1996, \apj, 465, 312,
  \dodoi{10.1086/177421}

\bibitem[{{Event Horizon Telescope Collaboration}
  {et~al.}(2019{\natexlab{a}}){Event Horizon Telescope Collaboration},
  {Akiyama}, {Alberdi}, {Alef}, {Asada}, {Azulay}, {Baczko}, {Ball},
  {Balokovi{\'c}}, {Barrett}, {Bintley}, {Blackburn}, {Boland}, {Bouman},
  {Bower}, {Bremer}, {Brinkerink}, {Brissenden}, {Britzen}, {Broderick},
  {Broguiere}, {Bronzwaer}, {Byun}, {Carlstrom}, {Chael}, {Chan}, {Chatterjee},
  {Chatterjee}, {Chen}, {Chen}, {Cho}, {Christian}, {Conway}, {Cordes}, {Crew},
  {Cui}, {Davelaar}, {De Laurentis}, {Deane}, {Dempsey}, {Desvignes}, {Dexter},
  {Doeleman}, {Eatough}, {Falcke}, {Fish}, {Fomalont}, {Fraga-Encinas},
  {Friberg}, {Fromm}, {G{\'o}mez}, {Galison}, {Gammie}, {Garc{\'\i}a},
  {Gentaz}, {Georgiev}, {Goddi}, {Gold}, {Gu}, {Gurwell}, {Hada}, {Hecht},
  {Hesper}, {Ho}, {Ho}, {Honma}, {Huang}, {Huang}, {Hughes}, {Ikeda}, {Inoue},
  {Issaoun}, {James}, {Jannuzi}, {Janssen}, {Jeter}, {Jiang}, {Johnson},
  {Jorstad}, {Jung}, {Karami}, {Karuppusamy}, {Kawashima}, {Keating},
  {Kettenis}, {Kim}, {Kim}, {Kim}, {Kino}, {Koay}, {Koch}, {Koyama}, {Kramer},
  {Kramer}, {Krichbaum}, {Kuo}, {Lauer}, {Lee}, {Li}, {Li}, {Lindqvist}, {Liu},
  {Liuzzo}, {Lo}, {Lobanov}, {Loinard}, {Lonsdale}, {Lu}, {MacDonald}, {Mao},
  {Markoff}, {Marrone}, {Marscher}, {Mart{\'\i}-Vidal}, {Matsushita},
  {Matthews}, {Medeiros}, {Menten}, {Mizuno}, {Mizuno}, {Moran}, {Moriyama},
  {Moscibrodzka}, {M{\"u}ller}, {Nagai}, {Nagar}, {Nakamura}, {Narayan},
  {Narayanan}, {Natarajan}, {Neri}, {Ni}, {Noutsos}, {Okino}, {Olivares},
  {Oyama}, {{\"O}zel}, {Palumbo}, {Patel}, {Pen}, {Pesce}, {Pi{\'e}tu},
  {Plambeck}, {PopStefanija}, {Porth}, {Prather}, {Preciado-L{\'o}pez},
  {Psaltis}, {Pu}, {Ramakrishnan}, {Rao}, {Rawlings}, {Raymond}, {Rezzolla},
  {Ripperda}, {Roelofs}, {Rogers}, {Ros}, {Rose}, {Roshanineshat}, {Rottmann},
  {Roy}, {Ruszczyk}, {Ryan}, {Rygl}, {S{\'a}nchez}, {S{\'a}nchez-Arguelles},
  {Sasada}, {Savolainen}, {Schloerb}, {Schuster}, {Shao}, {Shen}, {Small},
  {Sohn}, {SooHoo}, {Tazaki}, {Tiede}, {Tilanus}, {Titus}, {Toma}, {Torne},
  {Trent}, {Trippe}, {Tsuda}, {van Bemmel}, {van Langevelde}, {van Rossum},
  {Wagner}, {Wardle}, {Weintroub}, {Wex}, {Wharton}, {Wielgus}, {Wong}, {Wu},
  {Young}, {Young}, {Younsi}, {Yuan}, {Yuan}, {Zensus}, {Zhao}, {Zhao}, {Zhu},
  {Farah}, {Meyer-Zhao}, {Michalik}, {Nadolski}, {Nishioka}, {Pradel},
  {Primiani}, {Souccar}, {Vertatschitsch}, \& {Yamaguchi}}]{2017_Mass_EHT}
{Event Horizon Telescope Collaboration}, {Akiyama}, K., {Alberdi}, A., {et~al.}
  2019{\natexlab{a}}, \apjl, 875, L6, \dodoi{10.3847/2041-8213/ab1141}

\bibitem[{{Event Horizon Telescope Collaboration}
  {et~al.}(2019{\natexlab{b}}){Event Horizon Telescope Collaboration},
  {Akiyama}, {Alberdi}, {Alef}, {Asada}, {Azulay}, {Baczko}, {Ball},
  {Balokovi{\'c}}, {Barrett}, {Bintley}, {Blackburn}, {Boland}, {Bouman},
  {Bower}, {Bremer}, {Brinkerink}, {Brissenden}, {Britzen}, {Broderick},
  {Broguiere}, {Bronzwaer}, {Byun}, {Carlstrom}, {Chael}, {Chan}, {Chatterjee},
  {Chatterjee}, {Chen}, {Chen}, {Cho}, {Christian}, {Conway}, {Cordes}, {Crew},
  {Cui}, {Davelaar}, {De Laurentis}, {Deane}, {Dempsey}, {Desvignes}, {Dexter},
  {Doeleman}, {Eatough}, {Falcke}, {Fish}, {Fomalont}, {Fraga-Encinas},
  {Friberg}, {Fromm}, {G{\'o}mez}, {Galison}, {Gammie}, {Garc{\'\i}a},
  {Gentaz}, {Georgiev}, {Goddi}, {Gold}, {Gu}, {Gurwell}, {Hada}, {Hecht},
  {Hesper}, {Ho}, {Ho}, {Honma}, {Huang}, {Huang}, {Hughes}, {Ikeda}, {Inoue},
  {Issaoun}, {James}, {Jannuzi}, {Janssen}, {Jeter}, {Jiang}, {Johnson},
  {Jorstad}, {Jung}, {Karami}, {Karuppusamy}, {Kawashima}, {Keating},
  {Kettenis}, {Kim}, {Kim}, {Kim}, {Kino}, {Koay}, {Koch}, {Koyama}, {Kramer},
  {Kramer}, {Krichbaum}, {Kuo}, {Lauer}, {Lee}, {Li}, {Li}, {Lindqvist}, {Liu},
  {Liuzzo}, {Lo}, {Lobanov}, {Loinard}, {Lonsdale}, {Lu}, {MacDonald}, {Mao},
  {Markoff}, {Marrone}, {Marscher}, {Mart{\'\i}-Vidal}, {Matsushita},
  {Matthews}, {Medeiros}, {Menten}, {Mizuno}, {Mizuno}, {Moran}, {Moriyama},
  {Moscibrodzka}, {Mul{\ensuremath{\ddot{}}}ler}, {Nagai}, {Nagar}, {Nakamura},
  {Narayan}, {Narayanan}, {Natarajan}, {Neri}, {Ni}, {Noutsos}, {Okino},
  {Olivares}, {Oyama}, {{\"O}zel}, {Palumbo}, {Patel}, {Pen}, {Pesce},
  {Pi{\'e}tu}, {Plambeck}, {PopStefanija}, {Porth}, {Prather},
  {Preciado-L{\'o}pez}, {Psaltis}, {Pu}, {Ramakrishnan}, {Rao}, {Rawlings},
  {Raymond}, {Rezzolla}, {Ripperda}, {Roelofs}, {Rogers}, {Ros}, {Rose},
  {Roshanineshat}, {Rottmann}, {Roy}, {Ruszczyk}, {Ryan}, {Rygl},
  {S{\'a}nchez}, {S{\'a}nchez-Arguelles}, {Sasada}, {Savolainen}, {Schloerb},
  {Schuster}, {Shao}, {Shen}, {Small}, {Sohn}, {SooHoo}, {Tazaki}, {Tiede},
  {Tilanus}, {Titus}, {Toma}, {Torne}, {Trent}, {Trippe}, {Tsuda}, {van
  Bemmel}, {van Langevelde}, {van Rossum}, {Wagner}, {Wardle}, {Weintroub},
  {Wex}, {Wharton}, {Wielgus}, {Wong}, {Wu}, {Young}, {Young}, {Younsi},
  {Yuan}, {Yuan}, {Zensus}, {Zhao}, {Zhao}, {Zhu}, {Anczarski}, {Baganoff},
  {Eckart}, {Farah}, {Haggard}, {Meyer-Zhao}, {Michalik}, {Nadolski},
  {Neilsen}, {Nishioka}, {Nowak}, {Pradel}, {Primiani}, {Souccar},
  {Vertatschitsch}, {Yamaguchi}, \& {Zhang}}]{EHT2019_BZ_BP}
---. 2019{\natexlab{b}}, \apjl, 875, L5, \dodoi{10.3847/2041-8213/ab0f43}

\bibitem[{{Event Horizon Telescope Collaboration}
  {et~al.}(2019{\natexlab{c}}){Event Horizon Telescope Collaboration},
  {Akiyama}, {Alberdi}, {Alef}, {Asada}, {Azulay}, {Baczko}, {Ball},
  {Balokovi{\'c}}, {Barrett}, {Bintley}, {Blackburn}, {Boland}, {Bouman},
  {Bower}, {Bremer}, {Brinkerink}, {Brissenden}, {Britzen}, {Broderick},
  {Broguiere}, {Bronzwaer}, {Byun}, {Carlstrom}, {Chael}, {Chan}, {Chatterjee},
  {Chatterjee}, {Chen}, {Chen}, {Cho}, {Christian}, {Conway}, {Cordes}, {Crew},
  {Cui}, {Davelaar}, {De Laurentis}, {Deane}, {Dempsey}, {Desvignes}, {Dexter},
  {Doeleman}, {Eatough}, {Falcke}, {Fish}, {Fomalont}, {Fraga-Encinas},
  {Freeman}, {Friberg}, {Fromm}, {G{\'o}mez}, {Galison}, {Gammie},
  {Garc{\'\i}a}, {Gentaz}, {Georgiev}, {Goddi}, {Gold}, {Gu}, {Gurwell},
  {Hada}, {Hecht}, {Hesper}, {Ho}, {Ho}, {Honma}, {Huang}, {Huang}, {Hughes},
  {Ikeda}, {Inoue}, {Issaoun}, {James}, {Jannuzi}, {Janssen}, {Jeter}, {Jiang},
  {Johnson}, {Jorstad}, {Jung}, {Karami}, {Karuppusamy}, {Kawashima},
  {Keating}, {Kettenis}, {Kim}, {Kim}, {Kim}, {Kino}, {Koay}, {Koch}, {Koyama},
  {Kramer}, {Kramer}, {Krichbaum}, {Kuo}, {Lauer}, {Lee}, {Li}, {Li},
  {Lindqvist}, {Liu}, {Liuzzo}, {Lo}, {Lobanov}, {Loinard}, {Lonsdale}, {Lu},
  {MacDonald}, {Mao}, {Markoff}, {Marrone}, {Marscher}, {Mart{\'\i}-Vidal},
  {Matsushita}, {Matthews}, {Medeiros}, {Menten}, {Mizuno}, {Mizuno}, {Moran},
  {Moriyama}, {Moscibrodzka}, {M{\"u}ller}, {Nagai}, {Nagar}, {Nakamura},
  {Narayan}, {Narayanan}, {Natarajan}, {Neri}, {Ni}, {Noutsos}, {Okino},
  {Olivares}, {Ortiz-Le{\'o}n}, {Oyama}, {{\"O}zel}, {Palumbo}, {Patel}, {Pen},
  {Pesce}, {Pi{\'e}tu}, {Plambeck}, {PopStefanija}, {Porth}, {Prather},
  {Preciado-L{\'o}pez}, {Psaltis}, {Pu}, {Ramakrishnan}, {Rao}, {Rawlings},
  {Raymond}, {Rezzolla}, {Ripperda}, {Roelofs}, {Rogers}, {Ros}, {Rose},
  {Roshanineshat}, {Rottmann}, {Roy}, {Ruszczyk}, {Ryan}, {Rygl},
  {S{\'a}nchez}, {S{\'a}nchez-Arguelles}, {Sasada}, {Savolainen}, {Schloerb},
  {Schuster}, {Shao}, {Shen}, {Small}, {Sohn}, {SooHoo}, {Tazaki}, {Tiede},
  {Tilanus}, {Titus}, {Toma}, {Torne}, {Trent}, {Trippe}, {Tsuda}, {van
  Bemmel}, {van Langevelde}, {van Rossum}, {Wagner}, {Wardle}, {Weintroub},
  {Wex}, {Wharton}, {Wielgus}, {Wong}, {Wu}, {Young}, {Young}, {Younsi},
  {Yuan}, {Yuan}, {Zensus}, {Zhao}, {Zhao}, {Zhu}, {Algaba}, {Allardi},
  {Amestica}, {Anczarski}, {Bach}, {Baganoff}, {Beaudoin}, {Benson},
  {Berthold}, {Blanchard}, {Blundell}, {Bustamente}, {Cappallo},
  {Castillo-Dom{\'\i}nguez}, {Chang}, {Chang}, {Chang}, {Chen}, {Chilson},
  {Chuter}, {C{\'o}rdova Rosado}, {Coulson}, {Crawford}, {Crowley}, {David},
  {Derome}, {Dexter}, {Dornbusch}, {Dudevoir}, {Dzib}, {Eckart}, {Eckert},
  {Erickson}, {Everett}, {Faber}, {Farah}, {Fath}, {Folkers}, {Forbes},
  {Freund}, {G{\'o}mez-Ruiz}, {Gale}, {Gao}, {Geertsema}, {Graham}, {Greer},
  {Grosslein}, {Gueth}, {Haggard}, {Halverson}, {Han}, {Han}, {Hao},
  {Hasegawa}, {Henning}, {Hern{\'a}ndez-G{\'o}mez}, {Herrero-Illana},
  {Heyminck}, {Hirota}, {Hoge}, {Huang}, {Impellizzeri}, {Jiang}, {Kamble},
  {Keisler}, {Kimura}, {Kono}, {Kubo}, {Kuroda}, {Lacasse}, {Laing}, {Leitch},
  {Li}, {Lin}, {Liu}, {Liu}, {Lu}, {Marson}, {Martin-Cocher}, {Massingill},
  {Matulonis}, {McColl}, {McWhirter}, {Messias}, {Meyer-Zhao}, {Michalik},
  {Monta{\~n}a}, {Montgomerie}, {Mora-Klein}, {Muders}, {Nadolski}, {Navarro},
  {Neilsen}, {Nguyen}, {Nishioka}, {Norton}, {Nowak}, {Nystrom}, {Ogawa},
  {Oshiro}, {Oyama}, {Parsons}, {Paine}, {Pe{\~n}alver}, {Phillips}, {Poirier},
  {Pradel}, {Primiani}, {Raffin}, {Rahlin}, {Reiland}, {Risacher}, {Ruiz},
  {S{\'a}ez-Mada{\'\i}n}, {Sassella}, {Schellart}, {Shaw}, {Silva}, {Shiokawa},
  {Smith}, {Snow}, {Souccar}, {Sousa}, {Sridharan}, {Srinivasan}, {Stahm},
  {Stark}, {Story}, {Timmer}, {Vertatschitsch}, {Walther}, {Wei}, {Whitehorn},
  {Whitney}, {Woody}, {Wouterloot}, {Wright}, {Yamaguchi}, {Yu}, {Zeballos},
  {Zhang}, \& {Ziurys}}]{EHT2017A}
---. 2019{\natexlab{c}}, \apjl, 875, L1, \dodoi{10.3847/2041-8213/ab0ec7}

\bibitem[{{Event Horizon Telescope Collaboration} {et~al.}(2021){Event Horizon
  Telescope Collaboration}, {Akiyama}, {Algaba}, {Alberdi}, {Alef}, {Anantua},
  {Asada}, {Azulay}, {Baczko}, {Ball}, {Balokovi{\'c}}, {Barrett}, {Benson},
  {Bintley}, {Blackburn}, {Blundell}, {Boland}, {Bouman}, {Bower}, {Boyce},
  {Bremer}, {Brinkerink}, {Brissenden}, {Britzen}, {Broderick}, {Broguiere},
  {Bronzwaer}, {Byun}, {Carlstrom}, {Chael}, {Chan}, {Chatterjee},
  {Chatterjee}, {Chen}, {Chen}, {Chesler}, {Cho}, {Christian}, {Conway},
  {Cordes}, {Crawford}, {Crew}, {Cruz-Osorio}, {Cui}, {Davelaar}, {De
  Laurentis}, {Deane}, {Dempsey}, {Desvignes}, {Dexter}, {Doeleman}, {Eatough},
  {Falcke}, {Farah}, {Fish}, {Fomalont}, {Ford}, {Fraga-Encinas}, {Friberg},
  {Fromm}, {Fuentes}, {Galison}, {Gammie}, {Garc{\'\i}a}, {Gelles}, {Gentaz},
  {Georgiev}, {Goddi}, {Gold}, {G{\'o}mez}, {G{\'o}mez-Ruiz}, {Gu}, {Gurwell},
  {Hada}, {Haggard}, {Hecht}, {Hesper}, {Himwich}, {Ho}, {Ho}, {Honma},
  {Huang}, {Huang}, {Hughes}, {Ikeda}, {Inoue}, {Issaoun}, {James}, {Jannuzi},
  {Janssen}, {Jeter}, {Jiang}, {Jimenez-Rosales}, {Johnson}, {Jorstad}, {Jung},
  {Karami}, {Karuppusamy}, {Kawashima}, {Keating}, {Kettenis}, {Kim}, {Kim},
  {Kim}, {Kim}, {Kino}, {Koay}, {Kofuji}, {Koch}, {Koyama}, {Kramer}, {Kramer},
  {Krichbaum}, {Kuo}, {Lauer}, {Lee}, {Levis}, {Li}, {Li}, {Lindqvist}, {Lico},
  {Lindahl}, {Liu}, {Liu}, {Liuzzo}, {Lo}, {Lobanov}, {Loinard}, {Lonsdale},
  {Lu}, {MacDonald}, {Mao}, {Marchili}, {Markoff}, {Marrone}, {Marscher},
  {Mart{\'\i}-Vidal}, {Matsushita}, {Matthews}, {Medeiros}, {Menten}, {Mizuno},
  {Mizuno}, {Moran}, {Moriyama}, {Moscibrodzka}, {M{\"u}ller}, {Musoke}, {Mus
  Mej{\'\i}as}, {Michalik}, {Nadolski}, {Nagai}, {Nagar}, {Nakamura},
  {Narayan}, {Narayanan}, {Natarajan}, {Nathanail}, {Neilsen}, {Neri}, {Ni},
  {Noutsos}, {Nowak}, {Okino}, {Olivares}, {Ortiz-Le{\'o}n}, {Oyama},
  {{\"O}zel}, {Palumbo}, {Park}, {Patel}, {Pen}, {Pesce}, {Pi{\'e}tu},
  {Plambeck}, {PopStefanija}, {Porth}, {P{\"o}tzl}, {Prather},
  {Preciado-L{\'o}pez}, {Psaltis}, {Pu}, {Ramakrishnan}, {Rao}, {Rawlings},
  {Raymond}, {Rezzolla}, {Ricarte}, {Ripperda}, {Roelofs}, {Rogers}, {Ros},
  {Rose}, {Roshanineshat}, {Rottmann}, {Roy}, {Ruszczyk}, {Rygl},
  {S{\'a}nchez}, {S{\'a}nchez-Arguelles}, {Sasada}, {Savolainen}, {Schloerb},
  {Schuster}, {Shao}, {Shen}, {Small}, {Sohn}, {SooHoo}, {Sun}, {Tazaki},
  {Tetarenko}, {Tiede}, {Tilanus}, {Titus}, {Toma}, {Torne}, {Trent},
  {Traianou}, {Trippe}, {van Bemmel}, {van Langevelde}, {van Rossum}, {Wagner},
  {Ward-Thompson}, {Wardle}, {Weintroub}, {Wex}, {Wharton}, {Wielgus}, {Wong},
  {Wu}, {Yoon}, {Young}, {Young}, {Younsi}, {Yuan}, {Yuan}, {Zensus}, {Zhao},
  \& {Zhao}}]{MagFieldEHT}
{Event Horizon Telescope Collaboration}, {Akiyama}, K., {Algaba}, J.~C.,
  {et~al.} 2021, \apjl, 910, L13, \dodoi{10.3847/2041-8213/abe4de}

\bibitem[{{Feng} \& {Wu}(2017)}]{Feng_accretion_jet_model}
{Feng}, J., \& {Wu}, Q. 2017, \mnras, 470, 612, \dodoi{10.1093/mnras/stx1283}

\bibitem[{{Franceschini} {et~al.}(2008){Franceschini}, {Rodighiero}, \&
  {Vaccari}}]{EBL_franceschini}
{Franceschini}, A., {Rodighiero}, G., \& {Vaccari}, M. 2008, \aap, 487, 837,
  \dodoi{10.1051/0004-6361:200809691}

\bibitem[{{Gaisser}(2013)}]{uhecr_power_requirement}
{Gaisser}, T.~K. 2013, in European Physical Journal Web of Conferences,
  Vol.~53, European Physical Journal Web of Conferences, 01012,
  \dodoi{10.1051/epjconf/20135301012}

\bibitem[{{Georganopoulos} \& {Kazanas}(2003)}]{structuredjet}
{Georganopoulos}, M., \& {Kazanas}, D. 2003, \apjl, 594, L27,
  \dodoi{10.1086/378557}

\bibitem[{{Ghisellini} {et~al.}(2005){Ghisellini}, {Tavecchio}, \&
  {Chiaberge}}]{SpineSheath_jets}
{Ghisellini}, G., {Tavecchio}, F., \& {Chiaberge}, M. 2005, \aap, 432, 401,
  \dodoi{10.1051/0004-6361:20041404}

\bibitem[{{Gilmore} {et~al.}(2012){Gilmore}, {Somerville}, {Primack}, \&
  {Dom{\'\i}nguez}}]{EBL_gilmore}
{Gilmore}, R.~C., {Somerville}, R.~S., {Primack}, J.~R., \& {Dom{\'\i}nguez},
  A. 2012, \mnras, 422, 3189, \dodoi{10.1111/j.1365-2966.2012.20841.x}

\bibitem[{{Giovannini} {et~al.}(1999){Giovannini}, {Taylor}, {Arbizzani},
  {Bondi}, {Cotton}, {Feretti}, {Lara}, \& {Venturi}}]{RG_structure}
{Giovannini}, G., {Taylor}, G.~B., {Arbizzani}, E., {et~al.} 1999, \apj, 522,
  101, \dodoi{10.1086/307640}

\bibitem[{{Giroletti} {et~al.}(2004){Giroletti}, {Giovannini}, {Feretti},
  {Cotton}, {Edwards}, {Lara}, {Marscher}, {Mattox}, {Piner}, \&
  {Venturi}}]{Mrk_structure}
{Giroletti}, M., {Giovannini}, G., {Feretti}, L., {et~al.} 2004, \apj, 600,
  127, \dodoi{10.1086/379663}

\bibitem[{{Hillas}(1984)}]{Hillas}
{Hillas}, A.~M. 1984, \araa, 22, 425,
  \dodoi{10.1146/annurev.aa.22.090184.002233}

\bibitem[{{Ho}(2008)}]{Review_DT_BLR}
{Ho}, L.~C. 2008, \araa, 46, 475,
  \dodoi{10.1146/annurev.astro.45.051806.110546}

\bibitem[{Hunter(2007)}]{matplotlib}
Hunter, J.~D. 2007, Computing In Science \& Engineering, 9,
  \dodoi{10.1109/MCSE.2007.55}

\bibitem[{{Ichimaru}(1977)}]{first_adaf}
{Ichimaru}, S. 1977, \apj, 214, 840, \dodoi{10.1086/155314}

\bibitem[{Jeff {et~al.}(2022)Jeff, {jbrockmendel}, Wes, Joris, Tom, Matthew,
  Simon, Phillip, {gfyoung}, {Sinhrks}, Patrick, Adam, Terji, Jeff, Chang,
  William, Shahar, JHM, Marc, Richard, Jeremy, Andy, Daniel, Edward, Fangchen,
  Matthew, Vytautas, Ali, Torsten, \& Pietro}]{panda_software}
Jeff, R., {jbrockmendel}, Wes, M., {et~al.} 2022, pandas-dev/pandas: Pandas
  1.4.2, v1.4.2,  Zenodo, \dodoi{10.5281/zenodo.6408044}

\bibitem[{{Kimura} {et~al.}(2015){Kimura}, {Murase}, \&
  {Toma}}]{ADAF_neutrinos_CR}
{Kimura}, S.~S., {Murase}, K., \& {Toma}, K. 2015, \apj, 806, 159,
  \dodoi{10.1088/0004-637X/806/2/159}

\bibitem[{{Kino} {et~al.}(2014){Kino}, {Takahara}, {Hada}, \& {Doi}}]{Kino_SSA}
{Kino}, M., {Takahara}, F., {Hada}, K., \& {Doi}, A. 2014, \apj, 786, 5,
  \dodoi{10.1088/0004-637X/786/1/5}

\bibitem[{{Kovalev}(2008)}]{M87_limb}
{Kovalev}, Y.~Y. 2008, in Astronomical Society of the Pacific Conference
  Series, Vol. 386, Extragalactic Jets: Theory and Observation from Radio to
  Gamma Ray, ed. T.~A. {Rector} \& D.~S. {De Young}, 155.
\newblock \doarXiv{0709.0953}

\bibitem[{{Kraichnan}(1965)}]{Kraichnan_turbulence}
{Kraichnan}, R.~H. 1965, Physics of Fluids, 8, 1385, \dodoi{10.1063/1.1761412}

\bibitem[{{Laor}(2003)}]{LLAGN_NLR}
{Laor}, A. 2003, \apj, 590, 86, \dodoi{10.1086/375008}

\bibitem[{{Mahadevan}(1997)}]{Mahadevan1997}
{Mahadevan}, R. 1997, \apj, 477, 585, \dodoi{10.1086/303727}

\bibitem[{{Manmoto} {et~al.}(1997){Manmoto}, {Mineshige}, \&
  {Kusunose}}]{ADAF_spectrum_Manmoto}
{Manmoto}, T., {Mineshige}, S., \& {Kusunose}, M. 1997, \apj, 489, 791,
  \dodoi{10.1086/304817}

\bibitem[{{Martin} {et~al.}(2019){Martin}, {Nixon}, {Pringle}, \&
  {Livio}}]{alpha_visco_obs}
{Martin}, R.~G., {Nixon}, C.~J., {Pringle}, J.~E., \& {Livio}, M. 2019, \na,
  70, 7, \dodoi{10.1016/j.newast.2019.01.001}

\bibitem[{{Massi}(2011)}]{SteadyReview}
{Massi}, M. 2011, \memsai, 82, 24.
\newblock \doarXiv{1010.3861}

\bibitem[{{Narayan} {et~al.}(2003){Narayan}, {Igumenshchev}, \&
  {Abramowicz}}]{MAD}
{Narayan}, R., {Igumenshchev}, I.~V., \& {Abramowicz}, M.~A. 2003, \pasj, 55,
  L69, \dodoi{10.1093/pasj/55.6.L69}

\bibitem[{{Narayan} {et~al.}(1997){Narayan}, {Kato}, \&
  {Honma}}]{Sonic_radius_critical}
{Narayan}, R., {Kato}, S., \& {Honma}, F. 1997, \apj, 476, 49,
  \dodoi{10.1086/303591}

\bibitem[{{Narayan} \& {Yi}(1995)}]{Narayan_Yi_original}
{Narayan}, R., \& {Yi}, I. 1995, \apj, 452, 710, \dodoi{10.1086/176343}

\bibitem[{{Nemmen} {et~al.}(2014){Nemmen}, {Storchi-Bergmann}, \&
  {Eracleous}}]{LLAGNmodels}
{Nemmen}, R.~S., {Storchi-Bergmann}, T., \& {Eracleous}, M. 2014, \mnras, 438,
  2804, \dodoi{10.1093/mnras/stt2388}

\bibitem[{P{\`e}rez \& {Granger}(2007)}]{ipython}
P{\`e}rez, F., \& {Granger}, B.~E. 2007, Computing in Science \& Engineering,
  9, 21, \dodoi{10.1109/MCSE.2007.53}

\bibitem[{{Perlman} {et~al.}(2007){Perlman}, {Mason}, {Packham}, {Levenson},
  {Elitzur}, {Schaefer}, {Imanishi}, {Sparks}, \& {Radomski}}]{No_DT_in_M87}
{Perlman}, E.~S., {Mason}, R.~E., {Packham}, C., {et~al.} 2007, \apj, 663, 808,
  \dodoi{10.1086/518781}

\bibitem[{{Prieto} {et~al.}(2016){Prieto}, {Fern{\'a}ndez-Ontiveros},
  {Markoff}, {Espada}, \& {Gonz{\'a}lez-Mart{\'\i}n}}]{Prieto_M87}
{Prieto}, M.~A., {Fern{\'a}ndez-Ontiveros}, J.~A., {Markoff}, S., {Espada}, D.,
  \& {Gonz{\'a}lez-Mart{\'\i}n}, O. 2016, \mnras, 457, 3801,
  \dodoi{10.1093/mnras/stw166}

\bibitem[{{Protheroe} {et~al.}(2003){Protheroe}, {Donea}, \&
  {Reimer}}]{M87UHECR_TeV}
{Protheroe}, R.~J., {Donea}, A.~C., \& {Reimer}, A. 2003, Astroparticle
  Physics, 19, 559, \dodoi{10.1016/S0927-6505(02)00268-2}

\bibitem[{{Protheroe} \& {Johnson}(1996)}]{Protheroe_Johnson_matrix}
{Protheroe}, R.~J., \& {Johnson}, P.~A. 1996, Astroparticle Physics, 4, 253,
  \dodoi{10.1016/0927-6505(95)00039-9}

\bibitem[{{Protheroe} \& {Stanev}(1993)}]{Protheroe_Stanev_matrix}
{Protheroe}, R.~J., \& {Stanev}, T. 1993, \mnras, 264, 191,
  \dodoi{10.1093/mnras/264.1.191}

\bibitem[{{Rees} {et~al.}(1982){Rees}, {Begelman}, {Blandford}, \&
  {Phinney}}]{first_adaf_torii}
{Rees}, M.~J., {Begelman}, M.~C., {Blandford}, R.~D., \& {Phinney}, E.~S. 1982,
  \nat, 295, 17, \dodoi{10.1038/295017a0}

\bibitem[{{Reimer} {et~al.}(2019){Reimer}, {B{\"o}ttcher}, \&
  {Buson}}]{Anita_matrix_intro}
{Reimer}, A., {B{\"o}ttcher}, M., \& {Buson}, S. 2019, \apj, 881, 46,
  \dodoi{10.3847/1538-4357/ab2bff}

\bibitem[{{Reimer} {et~al.}(2004){Reimer}, {Protheroe}, \& {Donea}}]{AnitaM87}
{Reimer}, A., {Protheroe}, R.~J., \& {Donea}, A.~C. 2004, \aap, 419, 89,
  \dodoi{10.1051/0004-6361:20034231}

\bibitem[{{Ressler} {et~al.}(2017){Ressler}, {Tchekhovskoy}, {Quataert}, \&
  {Gammie}}]{beta_values2}
{Ressler}, S.~M., {Tchekhovskoy}, A., {Quataert}, E., \& {Gammie}, C.~F. 2017,
  \mnras, 467, 3604, \dodoi{10.1093/mnras/stx364}

\bibitem[{{Shakura} \& {Sunyaev}(1973)}]{ShakuraSunyaev}
{Shakura}, N.~I., \& {Sunyaev}, R.~A. 1973, \aap, 500, 33

\bibitem[{{Sikora} {et~al.}(1997){Sikora}, {Madejski}, {Moderski}, \&
  {Poutanen}}]{dopplerSikora}
{Sikora}, M., {Madejski}, G., {Moderski}, R., \& {Poutanen}, J. 1997, \apj,
  484, 108, \dodoi{10.1086/304305}

\bibitem[{{Sol} {et~al.}(1989){Sol}, {Pelletier}, \& {Asseo}}]{Two-flow_jets}
{Sol}, H., {Pelletier}, G., \& {Asseo}, E. 1989, \mnras, 237, 411,
  \dodoi{10.1093/mnras/237.2.411}

\bibitem[{{Stawarz} {et~al.}(2006){Stawarz}, {Aharonian}, {Kataoka},
  {Ostrowski}, {Siemiginowska}, \& {Sikora}}]{Jet_power_Stawarz}
{Stawarz}, {\L}., {Aharonian}, F., {Kataoka}, J., {et~al.} 2006, \mnras, 370,
  981, \dodoi{10.1111/j.1365-2966.2006.10525.x}

\bibitem[{{Stawarz} {et~al.}(2003){Stawarz}, {Sikora}, \&
  {Ostrowski}}]{DopplerStawarz}
{Stawarz}, {\L}., {Sikora}, M., \& {Ostrowski}, M. 2003, \apj, 597, 186,
  \dodoi{10.1086/378290}

\bibitem[{{Stepney} \& {Guilbert}(1983)}]{rates_hotplasma}
{Stepney}, S., \& {Guilbert}, P.~W. 1983, \mnras, 204, 1269,
  \dodoi{10.1093/mnras/204.4.1269}

\bibitem[{{Svensson}(1982)}]{Svensson_brems}
{Svensson}, R. 1982, \apj, 258, 335, \dodoi{10.1086/160082}

\bibitem[{{Tavecchio} \& {Ghisellini}(2008)}]{M87_spine_sheath}
{Tavecchio}, F., \& {Ghisellini}, G. 2008, \mnras, 385, L98,
  \dodoi{10.1111/j.1745-3933.2008.00441.x}

\bibitem[{{van der Walt} {et~al.}(2011){van der Walt}, {Colbert}, \&
  {Varoquaux}}]{numpy}
{van der Walt}, S., {Colbert}, C.~S., \& {Varoquaux}, G. 2011, Computing in
  Science \& Engineering, 13, 22, \dodoi{10.1109/MCSE.2011.37}

\bibitem[{{W}es {M}c{K}inney(2010)}]{pandas}
{W}es {M}c{K}inney. 2010, in {P}roceedings of the 9th {P}ython in {S}cience
  {C}onference, ed. {S}t\'efan van~der {W}alt \& {J}arrod {M}illman, 56 -- 61,
  \dodoi{10.25080/Majora-92bf1922-00a}

\end{thebibliography}
\bibliographystyle{aasjournal}

\end{document}